\documentclass[11pt]{article}

\PassOptionsToPackage{table}{xcolor}
\usepackage[final,nohyperref]{acl}

\usepackage{times}
\usepackage{latexsym}
\usepackage{booktabs} 
\usepackage{graphicx}
\usepackage{tabularx}   
\usepackage{multirow}   
\usepackage{xcolor}     
\usepackage{pifont}
\usepackage{amsmath}
\usepackage{makecell} 
\usepackage{tcolorbox}
\usepackage[T1]{fontenc}

\usepackage[utf8]{inputenc}

\usepackage{microtype}

\usepackage{inconsolata}
\usepackage{subcaption}

\usepackage{graphicx}
\usepackage{arydshln} 

\usepackage{fontawesome5}

\newcommand{\cmark}{\textcolor{green}{\faCheck}}
\newcommand{\xmark}{\textcolor{red}{\faTimes}}
\usepackage{threeparttable}
\usepackage{listings}

\title{SpeechLLM-as-Judges: Towards General and Interpretable \\
Speech Quality Evaluation}

\author{%
    \textbf{Hui Wang}$^1$\thanks{Work done during an internship at Microsoft Research Asia.} \ \ 
    \textbf{Jinghua Zhao}$^1$ \ \
    \textbf{Yifan Yang}$^2$ \ \
    \textbf{Shujie Liu}$^2$\thanks{Corresponding authors.} \ \ 
    \textbf{Junyang Chen}$^1$ \ \ 
    \textbf{Yanzhe Zhang}$^1$ \\
    \textbf{Shiwan Zhao}$^1$ \ \
    \textbf{Jinyu Li}$^{2\dagger}$ \ \ 
    \textbf{Jiaming Zhou}$^1$ \ \
    \textbf{Haoqin Sun}$^1$ \ \
    \textbf{Yan Lu}$^2$ \ \
    \textbf{Yong Qin}$^{1\dagger}$ \\
    $^1$College of Computer Science, Nankai University \\
    $^2$Microsoft Corporation
\\
 \small{
   \textbf{Correspondence:} 
   \href{wanghui\_hlt@mail.nankai.edu.cn}{wanghui\_hlt@mail.nankai.edu.cn}, 
   \href{mailto:qinyong@nankai.edu.cn}{qinyong@nankai.edu.cn}
 }
}

\begin{document}
\maketitle
\begin{abstract}
Generative speech technologies are progressing rapidly, but evaluating the perceptual quality of synthetic speech remains a core challenge. Existing methods typically rely on scalar scores or binary decisions, which lack interpretability and generalization across tasks and languages. We present SpeechLLM-as-Judges, a new paradigm for enabling large language models (LLMs) to conduct structured and explanation-based speech quality evaluation. To support this direction, we introduce SpeechEval, a large-scale dataset containing 32,207 multilingual speech clips and 128,754 annotations spanning four tasks: quality assessment, pairwise comparison, improvement suggestion, and deepfake detection. Based on this resource, we develop SQ-LLM, a speech-quality-aware LLM trained with chain-of-thought reasoning and reward optimization to improve capability. Experimental results show that SQ-LLM delivers strong performance across tasks and languages, revealing the potential of this paradigm for advancing speech quality evaluation. The relevant code, models, and data are publicly available at \url{https://github.com/NKU-HLT/SpeechLLM-as-Judges}.
\end{abstract}

\section{Introduction}

Recent advances in generative modeling, including neural text-to-speech (TTS)~\citep{chen2025neural,wang2025felle}, speech-to-speech translation~\citep{seamlessm4t}, and large-scale spoken dialogue systems~\citep{omnichat}, have profoundly transformed the field of human-computer interaction. Modern applications such as voice assistants and conversational AI increasingly rely on the ability to generate natural, intelligible, and high-quality speech. In this context, accurately assessing the perceptual quality of the generated speech is essential to ensure the reliability of the system and guide the development of the model~\citep{vinay2022evaluating, yang2025towards}. However, achieving such evaluation remains challenging due to the complexity of perceptual quality factors, the diversity of speech generation tasks, and the lack of standardized, interpretable, and scalable assessment methods.

One of the primary challenges is the lack of interpretability in existing methods. Standard evaluation protocols such as Mean Opinion Score (MOS) and AB preference tests typically provide scalar scores or categorical judgments that reflect overall quality perception, but fail to offer insights into the specific factors that influence those judgments. Similarly, commonly used objective metrics such as Mel Cepstral Distortion (MCD)~\citep{kubichek1993mel} are designed to approximate signal similarity but not offer clear explanations related to human perception~\citep{vinay2022evaluating}. The absence of aspect-level attribution and structured reasoning limits their utility for targeted improvement and quality control, which in turn constrains the development of speech generation systems.

\begin{figure*}[t]
\begin{center}
\includegraphics[width=\linewidth]{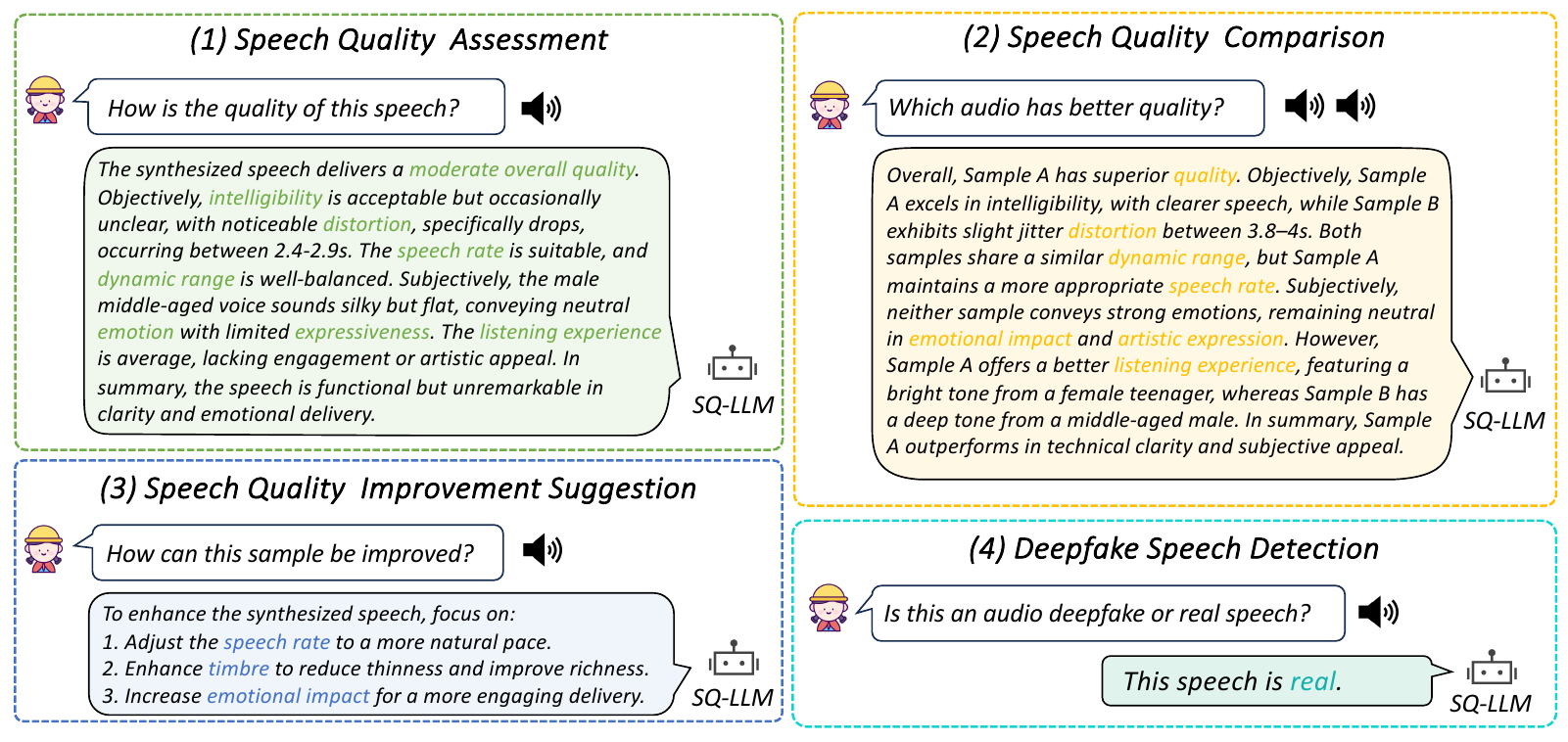}
\end{center}
\vspace{-10pt}
\caption{Example interactions showcasing the core capabilities of SpeechLLM-as-Judges. The model supports: speech quality assessment, comparison, improvement suggestion, and deepfake detection.}
\label{fig:model_case}
\end{figure*}

Another limitation lies in their limited ability to generalize across both data and task dimensions. On the data side, existing models are typically trained on narrowly scoped datasets, often restricted to a single language, a limited range of sources, and relatively simple annotations. Such constraints lead to weak robustness when models are deployed in multilingual, cross-domain, or real-world scenarios~\citep{RAMP, wang24s_interspeech}. On the task side, most evaluation frameworks are designed to perform a single function, such as absolute quality scoring or preference classification, and thus cannot simultaneously address other critical objectives, including generating improvement suggestions or detecting synthetic speech. Although recent speech foundation models provide architectural flexibility for multi-task evaluation~\citep{chu2024qwen20audio,dinkel2025midashenglm0}, they frequently underperform in quality assessment, primarily due to insufficient fine-grained, low-level perceptual reasoning~\citep{qualispeech}.

To address these challenges, we propose \textbf{SpeechLLM-as-Judges}, a new paradigm that enables large language models (LLMs) to perform general-purpose, interpretable speech quality evaluation. This approach equips existing speech large language models with the ability to reason about, explain, and compare perceptual aspects of speech quality across diverse tasks, as illustrated in Figure~\ref{fig:model_case}. To support this goal, we construct SpeechEval, a large-scale multilingual dataset comprising over 30,000 speech clips and 128,000 annotations, spanning four languages and covering multiple evaluation tasks, including descriptive speech quality assessment, pairwise comparison, quality-improvement suggestion, and deepfake speech detection. Building on SpeechEval, we design a speech-quality–aware large language model using instruction tuning with Chain-of-Thought (CoT) reasoning and reward optimization by Policy Gradient Optimization (GRPO) learning to enhance its capabilities. Comprehensive experiments not only validate the effectiveness and feasibility of this paradigm but also highlight its promising potential for advancing speech quality evaluation.

Our work makes the following key contributions:
\begin{itemize}
  \setlength{\itemsep}{0pt}
  \setlength{\parskip}{0pt}
  \setlength{\parsep}{0pt}
  \setlength{\topsep}{0pt}
   \item We present SpeechEval, a large-scale dataset for speech quality evaluation, containing 128,754 annotations across languages, domains, and evaluation types.
   \item We propose SQ-LLM, a speech-quality-aware model trained in two stages: instruction tuning with chain-of-thought reasoning and reward learning via GRPO. This approach enables the model to perform interpretable and general-purpose quality evaluation.
   \item We conduct extensive experiments to evaluate the effectiveness of the SpeechLLM-as-Judges paradigm, demonstrating its advantages in accuracy, interpretability, and generalization across tasks and domains.
\end{itemize}

\begin{table*}[ht]
\centering
\small
\setlength{\tabcolsep}{4pt} 
\resizebox{1\textwidth}{!}{%
\begin{threeparttable}
\begin{tabular}{@{}l c c c c c c c c}
    \midrule
    \multirow{2.5}{*}{\textbf{\makecell{Dataset}}} & \multirow{2.5}{*}{\textbf{\makecell{\# Label}}} & \multirow{2.5}{*}{\textbf{\makecell{Languages}}} & \multicolumn{2}{c}{\textbf{Quality Assessment}} & \multicolumn{2}{c}{\textbf{Quality Comparison}} & \multirow{2.5}{*}{\textbf{\makecell{Quality\\Suggestion}}} & \multirow{2.5}{*}{\textbf{\makecell{Deepfake\\Detection}}} \\
    \cmidrule(lr){4-5} \cmidrule(lr){6-7}
    & & & \textbf{Form} & \textbf{\# Dim} & \textbf{Form} & \textbf{\# Dim} & & \\
    \midrule
    ASVspoof2019-LA \citep{wang2020asvspoof} & 121{,}461 & English & \xmark & --- & \xmark & --- & \xmark & \cmark \\
    VCC2018 \citep{lorenzo2018voice} & 113{,}168$^{*}$ & English & Scores & 2 & \xmark & --- & \xmark & \xmark \\
    BC2019 \citep{wu2019blizzard} & 812 & Chinese & Scores & 1 & \xmark & --- & \xmark & \xmark \\
    BVCC \citep{cooper2021voices} & 7{,}106 & English & Scores & 1 & \xmark & --- & \xmark & \xmark \\
    NISQA \citep{mittag2021nisqa} & 14{,}672 & English & Scores & 5 & \xmark & --- & \xmark & \xmark \\
    QualiSpeech \citep{qualispeech} & 14{,}577 & English & \makecell{Natural-language\\\textcolor{gray}{(Human-annotated)}} & 7+4 & \xmark & --- & \xmark & \xmark \\
    ALLD-dataset \citep{chen2025audio} & 25{,}680 & English & \makecell{Natural-language\\\textcolor{gray}{(LLM-generated)}} & 5 & \makecell{Natural-language\\\textcolor{gray}{(LLM-generated)}} & 5 & \xmark & \xmark \\
    \midrule
    \textbf{SpeechEval(OURS)} & \textbf{128{,}754}  & \textbf{\makecell{Chinese, English,\\Japanese \& French}} & \textbf{\makecell{Natural-language\\\textcolor{gray}{(Human-annotated)}}} & \textbf{8+3+5} & \textbf{\makecell{Natural-language\\\textcolor{gray}{(Human-annotated)}}} & \textbf{8+3+5} & \cmark & \cmark \\
    \midrule
\end{tabular}

\begin{tablenotes}[flushleft]
\footnotesize
\item $^{*}$ The VCC2018 dataset comprises 113,168 human annotations, including 82,304 naturalness assessments and 30,864 speaker similarity assessments.
\end{tablenotes}
\end{threeparttable}
}
\vspace{-2pt}
\caption{Overview of existing speech quality evaluation datasets and task coverage.}
\vspace{-5pt}
\label{tab:dataset_overview}
\end{table*}

\section{Related Work}
\paragraph{Conventional Speech Quality Evaluation} 
Traditional methods for evaluating speech quality can be broadly categorized into subjective and objective approaches. Subjective protocols rely on human judgment and typically yield scalar or categorical outcomes. While these methods offer valuable perceptual insights, they are inherently time-consuming and labor-intensive~\citep{liu2025musiceval}. In contrast, objective metrics like the STOI and MCD~\citep{kubichek1993mel} aim to approximate perceptual quality through signal-based heuristics or learned mappings. Although these techniques are efficient, reproducible, and scalable, they often fall short in providing human-aligned explanations for their assessments~\citep{vinay2022evaluating}. These limitations collectively hinder the scalability, efficiency, and accuracy of speech quality evaluation in practical scenarios.

\paragraph{Task-Specific Models for Speech Quality Prediction}
Recent efforts have explored learning-based approaches for automatic speech quality prediction, including models such as MOSNet~\citep{MOSNet}, UTMOS~\citep{UTMOS}, Audiobox Aesthetics~\citep{tjandra2025meta}, and RAMP~\citep{RAMP,wang2025ramp+}. These systems are typically trained to regress MOS scores or classify quality labels. While such models improve over traditional metrics in terms of data-driven performance, they remain task-specific, language-constrained, and limited to scalar output formats. Few of them support multi-task inference or offer interpretable rationales for their predictions. As a result, their utility for generative speech workflows remains limited.

\begin{figure}[t]
\begin{center}
\includegraphics[width=\linewidth]{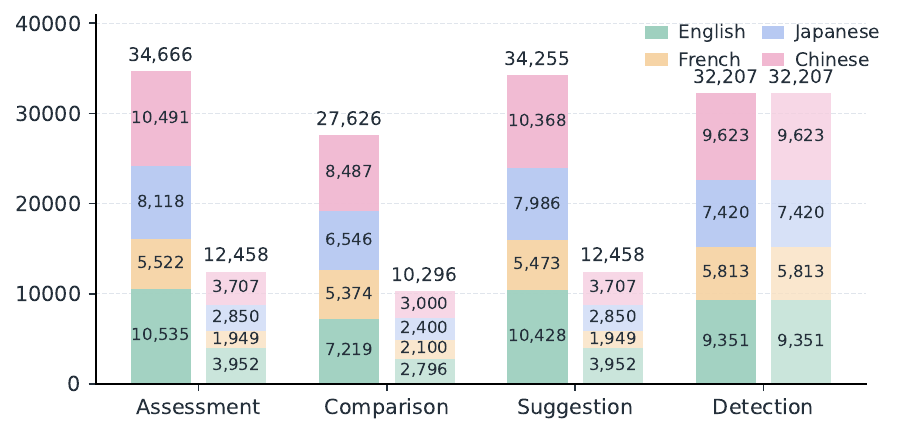}
\end{center}
\vspace{-6pt}
\caption{Per-task statistics across four languages, highlighting both the number of audio items (light color) and annotations (solid color).}
\label{fig:data_task_language}
\vspace{-2pt}
\end{figure}

\paragraph{LLMs for Speech Quality Understanding}
Recent advances in LLMs have motivated extending evaluation beyond text to speech~\citep{zhang2025speechjudge}. However, prior studies find that speech LLMs perform poorly on speech-quality benchmarks in direct evaluation and typically require task-specific tuning to produce reliable feedback~\citep{chen2025audio,qualispeech,10890590}. Nevertheless, as illustrated in Table~\ref{tab:dataset_overview}, their effectiveness remains limited by several factors: most models operate only on English, rely on semi-automatically generated or constrained datasets, and support only a narrow range of tasks. These limitations suggest that current speech LLMs have not yet acquired generalized quality understanding capabilities. In contrast, our work equips speech LLMs with multi-task capabilities through supervised training on large-scale data, enabling structured and interpretable judgments across languages, domains, and evaluation tasks.

\section{SpeechEval Dataset}

\begin{figure*}[h]
\begin{center}
\includegraphics[width=\linewidth]{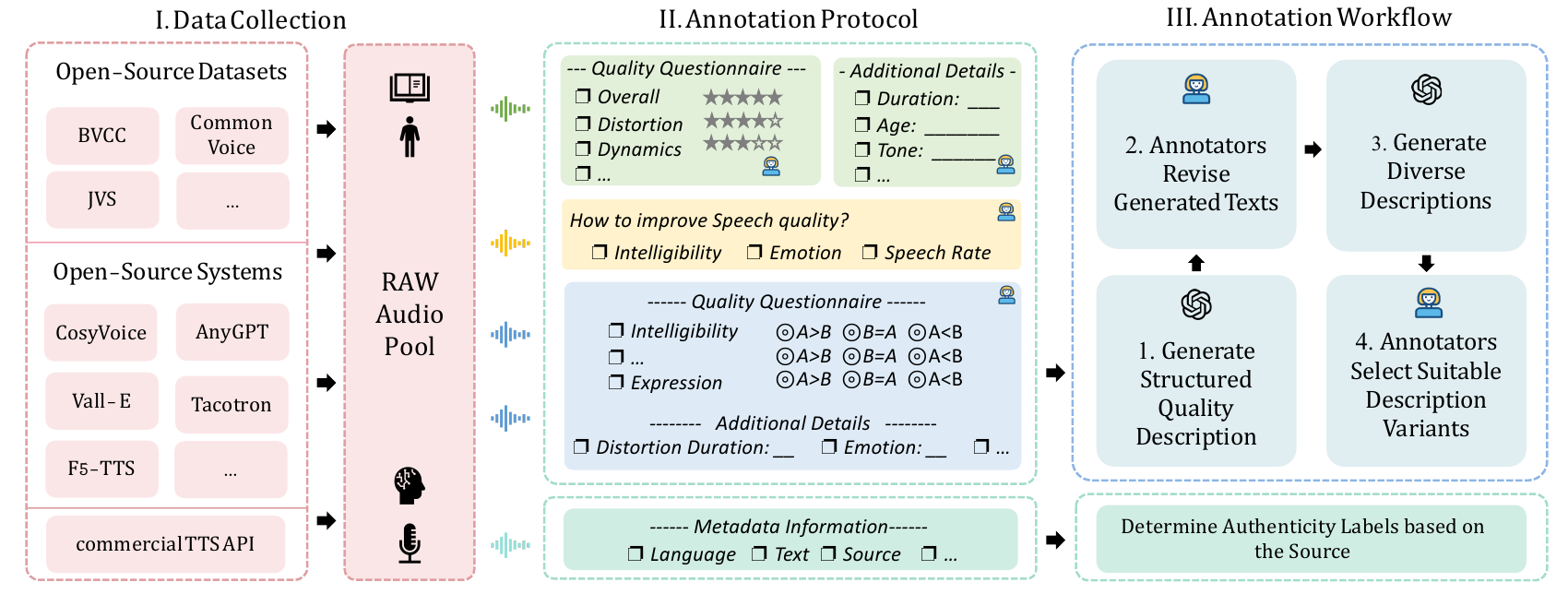}
\end{center}
\caption{Overview of the SpeechEval data construction process, including data collection (left), task-specific annotation protocols (middle), and a human-in-the-loop annotation workflow with LLM assistance (right).}
\label{fig:data construction}
\end{figure*}

\subsection{Task Formulation}
To move beyond simple scalar scores and limited task types, we propose four tasks that cover key real-world needs. 
The first task, \textbf{Speech Quality Assessment (SQA)}, focuses on generating natural language descriptions for a single utterance. The model is expected to articulate perceptual impressions across multiple dimensions. The second task, \textbf{Speech Quality Comparison (SQC)}, requires the model to compare two utterances and determine which one is of higher quality, along with a justification. This aligns with practical evaluation workflows like A/B testing. The third task, \textbf{Speech Quality Improvement Suggestion (SQI)}, introduces a corrective angle: given a suboptimal utterance, the model must suggest actionable modifications that could improve its quality. Lastly, \textbf{Deepfake Speech Detection (DSD)} concerns the ability to distinguish human speech from synthetic or manipulated speech. This task is closely tied to perceptual realism and quality degradation, making it a relevant subtask for quality-oriented systems.

\subsection{Dataset Overview}

The SpeechEval dataset contains 32,207 unique utterances and 128,754 annotations across four major tasks. Each task includes speech samples in English, Chinese, Japanese, and French, covering a broad range of speakers, speaking styles, voice qualities, and acoustic conditions. As shown in Figure~\ref{fig:data_task_language}, the data is distributed relatively evenly across tasks and languages. Chinese and English together account for the majority of samples, while Japanese and French are also well represented, ensuring multilingual coverage. All annotations are paired with structured labels and natural language explanations. A demonstration of the data, along with a detailed breakdown of the annotation and the associated metadata, can be found in Appendix~\ref{app:dataset}.

\subsection{Dataset Construction}

\paragraph{Data Collection}
We collect speech samples from multiple sources to ensure diversity in speaker characteristics, content, and quality. Specifically, we include real speech from public corpora~\citep{commonvoice:2020, lorenzo2018voice, cooper2021voices, panayotov2015librispeech, aishell3shi21c_interspeech, databaker, wu2019blizzard, sonobe2017jsut, takamichi2019jvs, honnet2017siwis}, along with synthetic speech generated using open-source speech generation systems~\citep{transformerli2019neural, vitskim2021conditional, tacotronwang17n_interspeech, fastspeech2chien2021investigating, zhan2024anygpt, zhang2023speechgpt, wang2025maskgct, Chen2024F5TTSAF, suno, chattts, du2024cosyvoice, chen2025neural}. Commercial TTS engines, such as Aliyun\footnote{Aliyun TTS: \url{https://ai.aliyun.com/nls/tts}}, Volcengine\footnote{Volcengine TTS: \url{https://console.volcengine.com}}, and Microsoft TTS\footnote{Microsoft TTS: \url{https://speech.microsoft.com/portal/voicegallery}}, are also included.

\begin{table}[t]
\centering
\renewcommand{\arraystretch}{1.02}
\begin{tabular}{p{2.2cm} p{4cm}}
\hline
\textbf{Aspects} & \textbf{Sub-dimensions} \\
\hline
\textbf{Overall} & (1) Overall Quality \\
\hdashline
\multirow[l]{4}{2.2cm}{\textbf{Production Quality}} & (2) Intelligibility \\
 & (3) Distortion \\
 & (4) Speech Rate \\
 & (5) Dynamic Range \\
\hdashline
\multirow[l]{3}{2.2cm}{\textbf{Content Enjoyment}} & (6) Emotional Impact \\
 & (7) Artistic Expression \\
 & (8) Subjective Experience \\
\hline
\end{tabular}
\caption{Structured speech quality annotation protocol. See Appendix~\ref{app:annotation protocol} for details.}
\vspace{-6pt}
\label{tab:annotation_protocol}
\end{table}

\begin{figure}[h]
\begin{center}
\includegraphics[width=\linewidth]{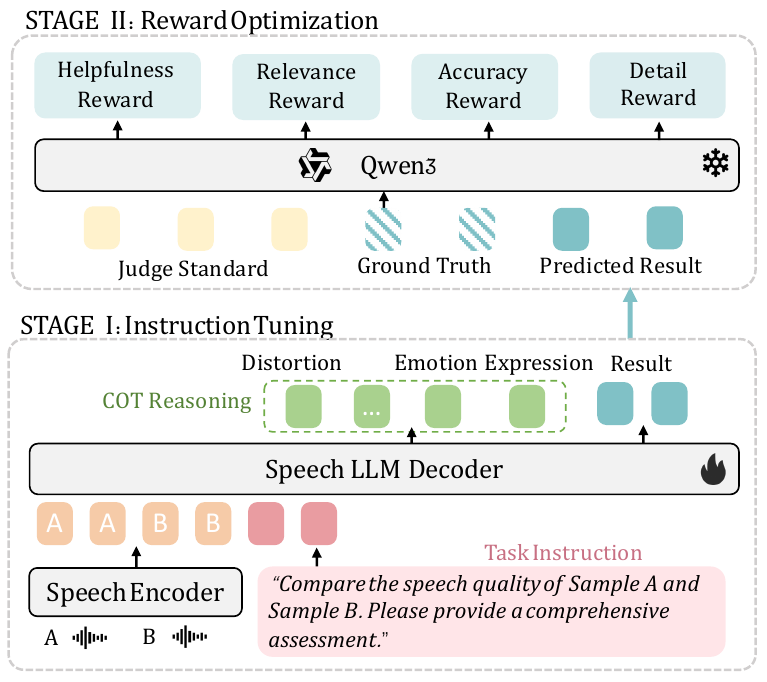}
\end{center}
\vspace{-6pt}
\caption{Overview of SQ-LLM training. Stage I uses instruction tuning with dimension-wise CoT reasoning. Stage II applies multi-aspect feedback for refinement.}
\vspace{-8pt}
\label{fig:model}
\end{figure}

\paragraph{Annotation Protocol} 
Each sample is annotated using a structured schema that captures speech quality. As shown in Table~\ref{tab:annotation_protocol}, we define three high-level aspects: overall rating, objective production quality, and subjective content enjoyment. These aspects are decomposed into eight subdimensions, including intelligence, distortion, speech rate, dynamic range, tone balance, emotional impact, artistic expression, and subjective experience. Depending on the task, annotators provide either ratings or pairwise comparisons for these dimensions. In addition, we collect three types of categorical metadata: distortion type, emotion type, and speaker gender. We also include five open-ended fields covering distortion duration, distortion severity, perceptual description, speaker age, and speaking tone. This annotation schema supports both structured supervision and natural language supervision.

\paragraph{Annotation Workflow}
To balance annotation quality with scalability, we adopt a human-in-the-loop pipeline as illustrated in Figure~\ref{fig:data construction}. For the Speech Quality Assessment, Comparison, and Suggestion tasks, annotators begin by completing a structured quality questionnaire that captures both judgments and additional information. The responses are passed to an LLM that generates initial textual descriptions. Annotators review and revise these drafts to ensure accuracy and clarity. The revised texts are subsequently passed back to the LLM to produce a set of diverse candidates, from which annotators select appropriate variants. This semi-automated process enables efficient production of high-quality and linguistically diverse annotations and is adaptable across both tasks and languages. For the Deepfake Speech Detection task, authenticity labels are derived based on metadata about the audio’s origin. These labels are assigned by tracing the source of the sample.

\section{SQ-LLM}

Figure~\ref{fig:model} presents SQ-LLM, a unified speech evaluation model. This section outlines its architecture and two-stage training; see Appendix~\ref{cotdata} for details.

\subsection{Model Architecture}

We design SQ-LLM to provide a unified solution for four distinct speech quality evaluation tasks: quality assessment, comparison, improvement suggestion, and deepfake detection. Rather than treating each task separately, SQ-LLM formulates them within a single instruction-based framework, where all tasks are handled through natural language prompts and structured outputs.

The model consists of a speech encoder and a speech-aware language decoder, both built on top of the Qwen2.5-Omni~\citep{xu2025qwen2050omni}. The encoder transforms input speech signals into continuous representations that capture relevant acoustic features. These are combined with the textual task instruction and passed to the decoder, which generates outputs such as quality dimension scores, free-form rationales, or classification results, depending on the task. This architecture allows SQ-LLM to generalize across tasks and languages while maintaining consistent, interpretable outputs.

\subsection{Instruction Tuning with CoT Reasoning}

To improve interpretability and consistency in generation, we adopt a reasoning-oriented instruction tuning strategy for the assessment, comparison, and suggestion tasks. Given a task instruction and one or more input utterances, the model first generates coarse predictions over a set of $N=8$ predefined quality dimensions. These dimension-wise predictions are used as intermediate signals that guide the final result. Since these annotations are already collected during human labeling, they serve as natural supervision targets for structured reasoning.

We define a joint training objective that encourages the model to produce accurate intermediate scores and coherent explanations. Formally, the overall loss is defined as:
\begin{equation}
\mathcal{L} = \lambda \sum_{i=1}^{N} \cdot \mathcal{L}_{\text{dim}}^{(i)} + \mathcal{L}_{\text{ans}},
\end{equation}
where $\mathcal{L}_{\text{dim}}^{(i)}$ denotes the loss for the $i$-th quality dimension, and $\mathcal{L}_{\text{ans}}$ is the cross-entropy loss for the final answer. $\lambda$ controls the contribution of intermediate reasoning.

\subsection{Reward Optimization via GRPO}

To further enhance the output alignment with human preferences, we apply Generalized Policy Gradient Optimization~\citep{deepseekai2025deepseekr1incentivizingreasoningcapability} to fine-tune SQ-LLM based on task-specific reward functions. Instead of relying on preference pairs, we design automatic reward evaluators for each task that score the model along four dimensions: Helpfulness, Relevance, Accuracy, and Level of Detail.

These rewards are produced by a single frozen evaluator, instantiated as Qwen3~\citep{yang2025qwen3}. Let \(p\) be the input prompt, \(y\) be the model output, \(g\) be the ground-truth metadata, and \(t \in \mathcal{T} = \{\text{Assessment}, \text{Comparison}, \text{Suggestion}\}\).
For each dimension, referred to as \(d \in \mathcal{D} = \{\text{Helpfulness},\text{Relevance},\text{Accuracy},\text{Detail}\}\), evaluator first returns an integer score  \(\mathcal{E}_d \in \{0, 1, 2 \ldots,10\}\), which we normalize and aggregate with fixed weights. We define normalized rewards \(r_d \in [0,1]\) and the total reward $R_{\text{total}}$:
\begin{equation}
\label{eq:reward-single}
r_d(p,y,g;t) =
\begin{cases}
\mathcal{E}_d(p,y,g)/10, & \text{if } t \in \mathcal{T}, \\[4pt]
\mathbf{1}\{y = g\}, & \text{if } t \notin \mathcal{T},
\end{cases}
\end{equation}

\begin{equation}
\label{eq:reward-total}
R_{\text{total}}
= \sum_{d \in \mathcal{D}} \lambda_d \, r_d.
\end{equation}

For instance, a quality analysis is rewarded when it accurately identifies the type and impact of distortion, while a comparative response is recognized for justifying its choice with coherent and well-reasoned arguments. The aggregated reward \(R_{\text{total}}\) serves as the learning signal within the GRPO framework to update the policy, improving output quality without sacrificing controllability.




\begin{table}[t]
\centering

\setlength{\tabcolsep}{4pt}
\renewcommand{\arraystretch}{1.08}
\begin{tabularx}{\linewidth}{l *{4}{>{\raggedleft\arraybackslash}X}}
\hline
\textbf{Task} & \textbf{Train} & \textbf{Val.} & \textbf{Test} & \textbf{Total} \\
\hline
Assessment 
  & 23,769 & 5,392  & 5,505  & 34,666 \\

Comparison 
  & 19,260 & 4,069  & 4,297  & 27,626 \\

Suggestion 
  & 23,494 & 5,316  & 5,445  & 34,255 \\

Detection
  & 6,600  & 5,724  & 19,883 & 32,207 \\
\hdashline
Total & 73,123  & 20,501  & 35,130 & 128,754\\
\hline
\end{tabularx}
\vspace{-5pt}
\caption{The data split across different tasks.}
\vspace{-5pt}
\label{tab:dataset_split}
\end{table}

\section{Experimental Setup}

\subsection{Implementation Details}

\paragraph{Dataset Split}
Table~\ref{tab:dataset_split} shows the data split. For the first three tasks, the audio data is divided into training, validation, and test sets in roughly a 70\%, 15\%, and 15\%. We ensure that validation and test splits contain unseen speakers, systems, and text content to evaluate generalization. 
For the detection task, we adopt a protocol based on prior work~\citep{wang2020asvspoof}. The dataset includes both seen and unseen spoof sources for in-distribution and zero-shot evaluation. The splits are imbalanced by design to reflect real-world distribution. Moreover, we remove cross-task data leakage. Full details of the splitting strategy are in Appendix~\ref{app:data_split}.

\paragraph{Model Configuration}
We fine-tune SQ-LLM based on Qwen2.5-Omni-7B using the Swift framework~\citep{zhao2025swift} with LoRA and the speech encoder frozen. The model is trained for 8 epochs, a batch size of 4, and a learning rate of 1e-4. and $\lambda = 0.3$ in the instruction tuning stage. We further apply GRPO with a batch size of 1, sampling 4 candidate generations per prompt, and train the LoRA adapters with a learning rate of 1e-6. $\lambda_{\text{Helpfulness}}$, $\lambda_{\text{Relevance}}$, $\lambda_{\text{Accuracy}}$ and $\lambda_{\text{Detail}}$ are set to 1, 1, 2 and 0.5, respectively. Overall, training SQ-LLM requires approximately 43 A100 GPU-hours, including about 12 GPU-hours for SFT and 31 GPU-hours for GRPO.


\begin{table*}[t]
\centering
\resizebox{0.95\linewidth}{!}{%
\renewcommand{\arraystretch}{1.25}
\begin{tabular}{lccccccccc} 
\hline
\multicolumn{1}{l}{\bf MODEL}  &
\multicolumn{3}{c}{\bf I. Quality Assessment} &
\multicolumn{3}{c}{\bf II. Quality Comparison} &
\multicolumn{3}{c}{\bf III. Improvement Suggestion} \\
\cmidrule(l{0.6em}r{0.6em}){2-4}
\cmidrule(l{0.6em}r{0.6em}){5-7}
\cmidrule(l{0.6em}r{0.6em}){8-10}
& {\bf LScore} & {\bf SBERT} & {\bf FENSE}
  & {\bf LScore} & {\bf SBERT} & {\bf FENSE}
  & {\bf LScore} & {\bf SBERT} & {\bf FENSE} \\
\hline
Qwen2-Audio-7B           & 4.161 & 0.580 & 0.311 & 4.591 & 0.703 & 0.580 & 5.176 & 0.548 & 0.460 \\
Qwen2.5-Omni-7B          & 4.610 & 0.593 & 0.593 & 3.602 & 0.704 & 0.437 & 6.426 & 0.537 & 0.364 \\
MiDashengLM-7B           & 5.536 & 0.699 & 0.635 & 4.042 & 0.665 & 0.444 & 6.653 & 0.600 & 0.490 \\
\hdashline
Qwen3-8B + Whisper       & 6.422 & \bf{0.803} & \bf{0.699} & 5.122 & 0.818 & 0.732 & \bf{7.478} & 0.583 & 0.574 \\
Qwen2.5-7B + AES-E       & 6.533 & 0.772 & 0.634 & 5.270 & 0.793 & 0.589 & 6.891 & 0.530 & 0.529 \\
Qwen3-4B + WavLM         & 6.163 & 0.736 & 0.585 & 5.138 & 0.796 & 0.618 & 7.209 & 0.539 & 0.537 \\
FT Qwen2-Audio-7B        & 6.440 & 0.800 & 0.559 & 5.648 & 0.853 & 0.686 & 7.263 & 0.708 & 0.708 \\
\hdashline
\textbf{SQ-LLM (OURS)}   & \bf{6.833} & \bf{0.803} & 0.614 & \bf{6.434} & \bf{0.864} & \bf{0.749} & 7.420 & \bf{0.735} & \bf{0.735} \\
\hline
\end{tabular}
}

\caption{Performance of different models on the Assessment, Comparison, and Improvement Suggestion tasks.}
\label{tab:unified_squ}
\end{table*}

\subsection{Metrics}
\label{sec:metrics}
For generation tasks (Assessment, Comparison, and Suggestion), we adopt a combination of automatic metrics, LLM-based, and human evaluation. Specifically, (i) the automatic metrics include SBERT and FENSE~\citep{zhang2019bertscore, zhou2022canfense}. (ii) We use LLM Score (LScore) to assess overall performance, and leverage an LLM to extract dimension-level predictions from generated texts: in the Assessment task, we report Pearson correlation coefficients (PCC) over 7 quality dimensions, together with accuracy (ACC) for speech-rate classification; in the Comparison task, we report ACC for dimension-wise preference. 
(iii) We conduct human evaluation to verify the models: we recruit $8$ raters to score model outputs on a 1--5 scale (higher is better). For each model, we sample $60$ instances, and each instance is independently rated by at least three raters.

For the Deepfake Speech Detection task, we report equal error rate (EER), minimum detection cost function (minDCF), and accuracy. The calculation details are given in Appendix~\ref{app:llmscore} and \ref{app:deepfakemetric}.

\subsection{Baselines}
\label{sec:baselines}

We consider three categories of baseline models to evaluate the effectiveness of SQ-LLM. First, we directly benchmark recent multimodal LLMs~\citep{chu2024qwen20audio,dinkel2025midashenglm0,xu2025qwen2050omni} without any task-specific tuning to assess their general capability for speech quality understanding. Second, we fine-tune existing open-source audio LLMs and our custom-constructed models by coupling advanced speech encoders (e.g., Whisper, WavLM, and AudioBox Aesthetics encoder (AES-E)~\citep{radford2023robust, chen2022wavlm, tjandra2025meta}) with an LLM backbone~\citep{yang2025qwen2}. Third, we include task-specific conventional expert models as strong non-LLM baselines; these systems are designed for a single task. Concretely, for the speech quality assessment and comparison tasks, we evaluate MOSNet, UTMOS, and AudioBox Aesthetics as representative expert baselines, while for the deepfake speech detection task, we adopt RawNet2, AASIST, and AASIST2. Detailed descriptions are provided in Appendix~\ref{app:baselines}.

\begin{table}[t]
\centering
\resizebox{0.95\linewidth}{!}{%
{%
\renewcommand{\arraystretch}{1.2}%
\begin{tabular}{lcc}
\hline
\multicolumn{1}{l}{\bf MODEL} &
\multicolumn{1}{c}{\bf Assessment} &
\multicolumn{1}{c}{\bf Comparison} \\
\hline
Qwen2-Audio-7B          & 0.073 & 0.310  \\
Qwen2.5-Omni-7B         & 0.070 & 0.203  \\
MiDashengLM-7B          & 0.158 & 0.265  \\
\hdashline
MOSNet      &0.333 & 0.726  \\
UTMOS   & 0.375 & 0.741 \\
Audiobox Aesthetics       & 0.464 & 0.737 \\
\hdashline
Qwen3-8B + Whisper      & 0.424 & 0.591  \\
Qwen2.5-7B + AES-E   & 0.457 & 0.577 \\
Qwen3-4B + WavLM        & 0.422 & 0.563 \\
FT Qwen2-Audio-7B       & 0.244 & 0.587 \\
\hdashline
\textbf{SQ-LLM (OURS)}  & \textbf{0.520} & \bf{0.751} \\
\hline
\end{tabular}%
}%
}
\vspace{-5pt}
\caption{Results report PCC for Assessment and ACC for Comparison, both on the Overall Quality dimension.
}
\vspace{-13pt}

\label{tab:traditionmos}
\end{table}

\begin{figure}[t]
\begin{center}
\includegraphics[width=\linewidth]{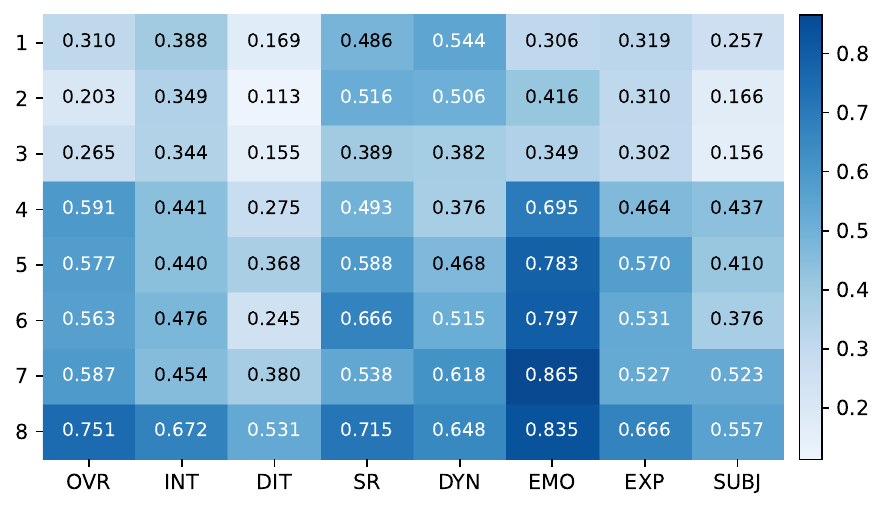}
\end{center}
\vspace{-10pt}
\caption{Heatmap of dimension-level accuracy across models on the Speech Quality Comparison task over eight quality dimensions. The 1–8 indices follow the same top-to-bottom order as in Table~\ref{tab:unified_squ}.}
\vspace{-5pt}
\label{fig:com-heatmap}
\end{figure}

\section{Results and Analysis}

\subsection{Comparison Studies}

\paragraph{Speech Quality Assessment} 
Table \ref{tab:unified_squ}-I presents the assessment results. In the direct-evaluation setting, models generally struggle to capture speech quality; MiDashengLM attains the highest LLM score among them, yet alignment with human ratings and text-based metrics remains weak. When we build custom-constructed pipelines or fine-tune audio LLMs, both the LLM score and the correlation with humans improve, accompanied by consistent gains on traditional metrics. SQ-LLM performs best overall, exhibiting the strongest agreement with human ratings and robust text metrics. 

Table~\ref{tab:traditionmos} shows that recent multimodal LLMs exhibit very low PCC on the Overall Quality dimension, indicating weak alignment with human. Expert MOS predictors markedly improve correlation, while SQ-LLM achieves the highest PCC, surpassing the best expert baseline and LLM variants.

\paragraph{Speech Quality Comparison}
On the comparison task (Table \ref{tab:unified_squ}-II), directly evaluated multimodal LLMs show limited ability to discriminate fine-grained quality differences: Qwen2-Audio is the strongest, whereas Qwen2.5-Omni and MiDashengLM lag behind. Equipping models with custom-constructed pipelines or fine-tuning audio LLMs delivers consistent gains. SQ-LLM further advances all dimensions, achieving the best agreement with human. From a correlation-oriented perspective in Table~\ref{tab:traditionmos}, models that better capture the monotonic quality ordering tend to yield higher comparison accuracy. Consistent with this trend, SQ-LLM attains the best ACC, outperforming expert baselines and LLM-based alternatives.

Figure~\ref{fig:com-heatmap} visualizes performance across eight quality dimensions. From a dimension-wise perspective, distortion and subjective experience are consistently the most challenging dimensions, with noticeably lower scores for most baselines, while speech rate and dynamic range are comparatively easier and show smaller performance gaps. Notably, SQ-LLM exhibits the most balanced strength, achieving clear gains on harder dimensions.

\paragraph{Speech Quality Improvement Suggestion}
Table \ref{tab:unified_squ}-III reports the results for generating actionable quality-improvement suggestions. Under direct evaluation, systems often produce vague or generic advice, MiDashengLM leads that group on the LLM score, but traditional metrics remain weak. Training markedly helps: both custom-constructed pipelines and fine-tuned audio LLMs improve lexical overlap and informativeness. Notably, Whisper encoder attains the strongest LLM score among the trained baselines. SQ-LLM offers the best overall balance of LLM scores and traditional metrics.

\begin{table}[t]

\centering
\resizebox{\linewidth}{!}{%
{%
\renewcommand{\arraystretch}{1.15}%
\begin{tabular}{lrrr}
\hline
\multicolumn{1}{l}{\bf MODEL} & \multicolumn{1}{c}{\bf EER(\%)} & \multicolumn{1}{c}{\bf minDCF} & \multicolumn{1}{c}{\bf ACC(\%)} \\
\hline
Qwen2-Audio-7B             & --     & --    & 21.220 \\
Qwen2.5-Omni-7B            & --     & --    & 27.250 \\
MiDashengLM-7B             & --     & --    & 67.480 \\
\hdashline
RawNet2                  & 15.836 & 0.325 & 72.036 \\
AASIST                   & 18.158 & 0.450 & 71.599 \\
AASIST2                  & 18.633 & 0.414 & 72.001 \\
\hdashline
Qwen3-8B + Whisper         & 16.502 & 0.411 & 78.977 \\
Qwen2.5-7B + AES-E         & 45.241 & 0.992 & 67.741 \\
Qwen3-4B + WavLM           & 44.300 & 1.000 & 20.837 \\
FT Qwen2-Audio-7B          & 8.593  & 0.194 & 89.312 \\
\hdashline
\textbf{SQ-LLM (OURS)}              & \textbf{6.249}  & \textbf{0.142} & \textbf{89.358} \\
\hline
\end{tabular}%
}%
\vspace{-4pt}
}
\caption{Results on deepfake speech detection.}

\label{tab:llm_eer_dcf_acc}
\end{table}

\begin{figure}[t]
\begin{center}
\includegraphics[width=\linewidth]{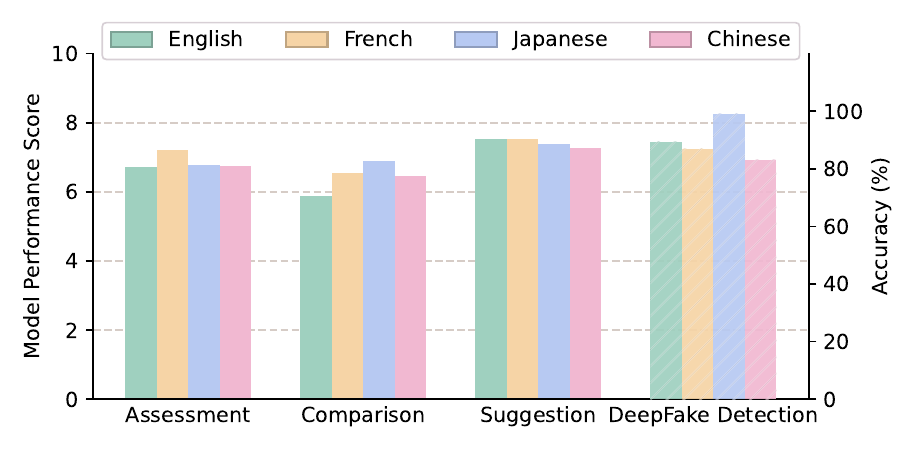}
\end{center}
\vspace{-10pt}
\caption{Performance of SQ-LLM across tasks and languages. Left y-axis reports LLM Scores; right y-axis shows accuracy for Deepfake Detection.}
\vspace{-10pt}
\label{fig:lang-task-results}
\end{figure}

\paragraph{Deepfake Speech Detection}
For detection task, results are summarized in Table \ref{tab:llm_eer_dcf_acc}. Expert systems provide solid, well-calibrated baselines. In contrast, untuned multimodal LLMs underperform and do not yield calibrated metrics, indicating weak out-of-the-box reliability. Custom-constructed model narrows the gap but is front-end sensitive: Whisper becomes competitive across all metrics, whereas other pipelines show marked degradation. Supervised fine-tuning helps substantially. While SQ-LLM achieves the best results overall, setting the strongest EER, minDCF, and accuracy.

\begin{figure}[t]
\begin{center}
\includegraphics[width=\linewidth]{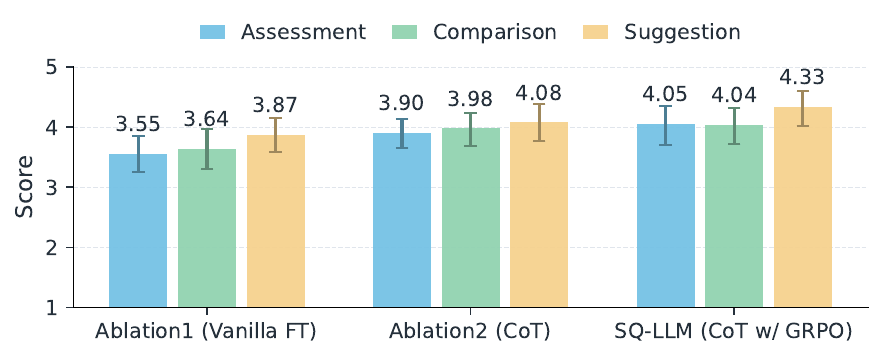}
\end{center}
\vspace{-10pt}
\caption{Human-evaluated ablation results on the Assessment, Comparison, and Suggestion tasks with 95\% item-level bootstrap confidence intervals.}
\label{fig:subj-task-results}
\end{figure}

\begin{table}[t]
\centering
\resizebox{0.90\linewidth}{!}{%
{%
\renewcommand{\arraystretch}{1.2}%
\begin{tabular}{cc|cccc}
\hline

\multicolumn{1}{c}{\bf CoT} &
\multicolumn{1}{c|}{\bf GRPO} &
\multicolumn{1}{c}{\bf SQA $\uparrow$} &
\multicolumn{1}{c}{\bf SQC $\uparrow$} &
\multicolumn{1}{c}{\bf SQI $\uparrow$} &
\multicolumn{1}{c}{\bf DSD $\downarrow$} \\
\hline
\cmark & \cmark & \textbf{6.833} & \textbf{6.434} & \textbf{7.420} & \textbf{6.249} \\
\cmark & \xmark                      & 6.804 & 6.420 & 7.018 & 6.264 \\
\xmark & \xmark                      & 6.657 & 6.316 & 6.733 & 8.574 \\
\hline
\end{tabular}%
}%
}

\caption{Ablation results across four tasks. Higher LLM Score is better for SQA, SQC, and SQI; EER (\%) is reported for DSD.}
\vspace{-15pt}
\label{tab:cot_grpo}
\centering
\end{table}

\subsection{Performance Analysis}

Figure~\ref{fig:lang-task-results} presents the performance of SQ-LLM across four tasks and four languages. The model demonstrates consistent and strong results in the Assessment, Comparison, and Suggestion tasks, with scores generally above 6.5 across all languages. For the Detection task, the model achieves high accuracy in Japanese, while performance in Chinese and French is relatively lower. These results highlight the model’s robust multilingual capability, and a more detailed analysis of the deepfake detection task is provided in Appendix~\ref{app:deepfake}.

In Figure~\ref{fig:subj-task-results} and Table~\ref{tab:cot_grpo}, we perform an ablation study to quantify the effects of CoT tuning and GRPO. Across both presentations, we observe consistent improvements as stronger components are introduced. Enabling CoT yields clear gains on the generation-style tasks, indicating that structured reasoning leads to more accurate and helpful outputs. Adding GRPO on top of CoT further improves performance, with the largest benefit on the Suggestion task, suggesting that preference optimization is particularly effective for open-ended recommendation generation. The deepfake detection results follow the same trend, where CoT substantially enhances robustness and the CoT-with-GRPO variant achieves the best overall performance. See Appendix~\ref{appendixresults} for a detailed analysis.

\section{Conclusion}
We present SpeechLLM-as-Judges, a unified framework for interpretable and generalizable speech quality evaluation. Supported by the SpeechEval dataset, SQ-LLM is trained to handle diverse evaluation tasks through instruction tuning with reasoning and reward optimization. Experiments demonstrate its ability to produce human-aligned, explanatory outputs. This work highlights the potential of speech LLM as a reliable and versatile evaluator in speech generation systems.


\section*{Limitations}
While our work provides a unified framework for speech quality evaluation, it also has limitations. The current version of SpeechEval focuses on four languages and a fixed set of tasks. Expanding to more low-resource or code-switched languages, as well as incorporating additional evaluation scenarios such as emotional expressiveness or speaker consistency, would further enhance the model’s coverage and applicability. We leave these directions for future work.

\section*{Ethics Statement}
All data used in this work were collected and processed in accordance with relevant ethical guidelines and licensing terms. The speech samples are sourced from publicly available or properly licensed datasets. Human annotations were conducted by trained annotators who received fair compensation. No personally identifiable information was collected during the annotation process. In this work, LLMs are employed for semi-automated annotation assistance and model evaluation, enhancing annotation efficiency and providing more accurate evaluation results.

\section*{Acknowledgement}
This work has been supported by the National Key R\&D Program of China (Grant No.2022ZD0116307) and NSF China (Grant No.62271270).

\bibliography{custom}


\appendix

\begin{figure}[htpb]
\begin{center}
\includegraphics[width=\linewidth]{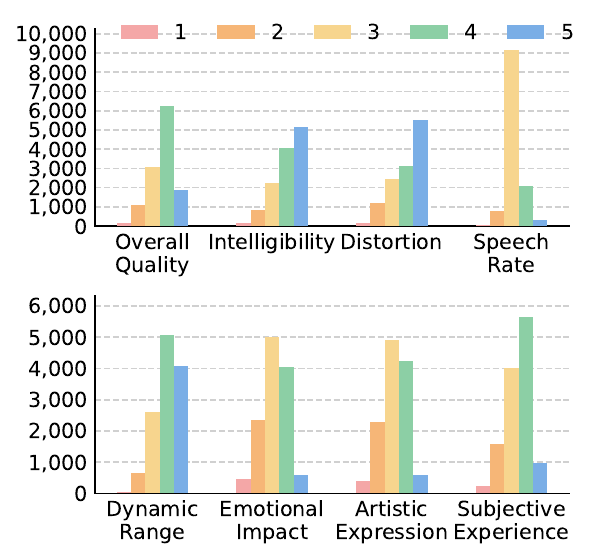}
\end{center}
\caption{Distribution of assessment scores across the eight dimensions. The speech rate dimension is rated from 1 (too slow) to 5 (too fast), while other dimensions use a 1–5 scale representing perceptual quality.}
\label{fig:score_distribution}
\end{figure}

\section{Dataset}
\label{app:dataset}

\subsection{Dataset Distribution}
\begin{figure}[t]
  \centering
  \begin{subfigure}{0.9\linewidth}
    \centering
    \includegraphics[width=\linewidth]{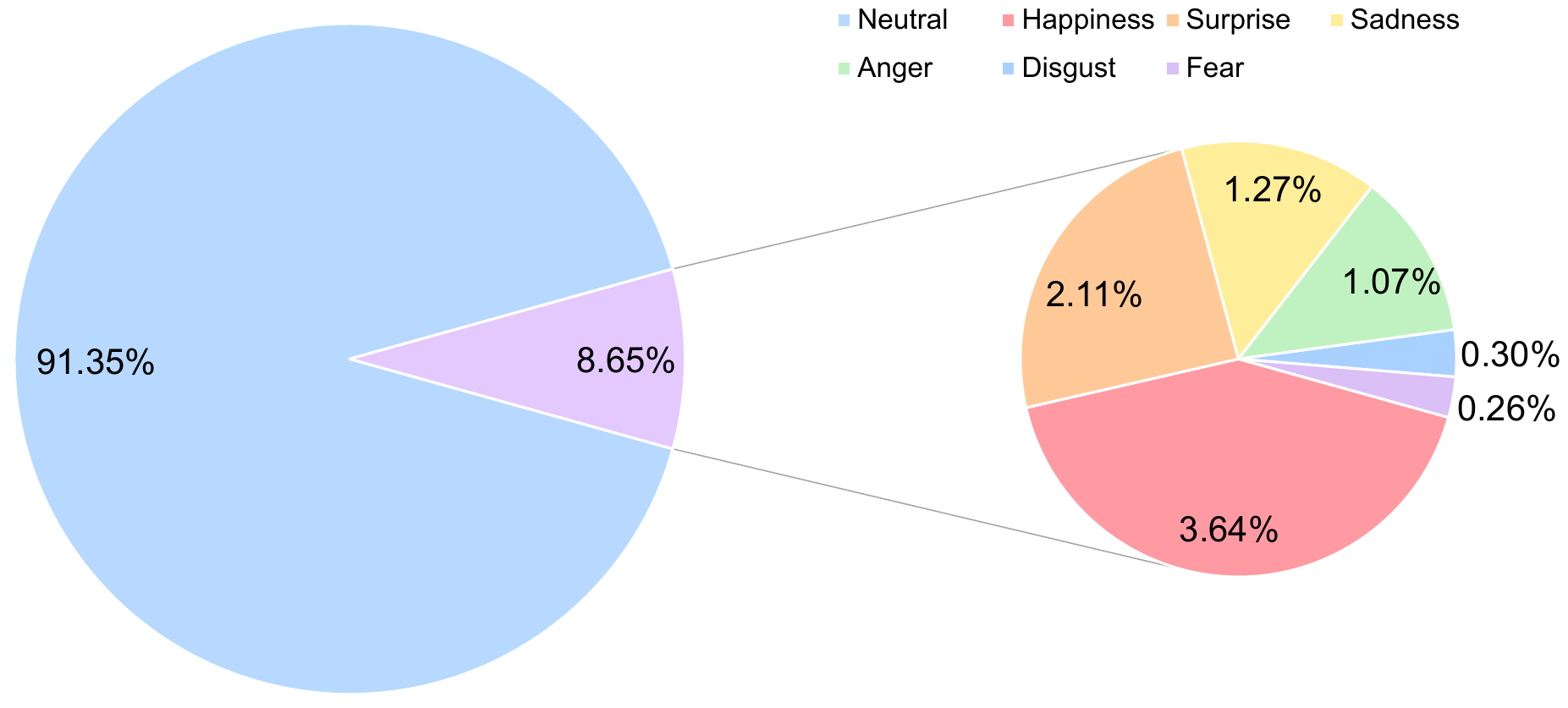}
    \caption{Emotion distribution: overall on the left, and the breakdown of non-neutral emotions on the right.}
    \label{fig:doublepie}
  \end{subfigure}

  \begin{subfigure}{0.42\linewidth}
    \centering
    \includegraphics[width=\linewidth]{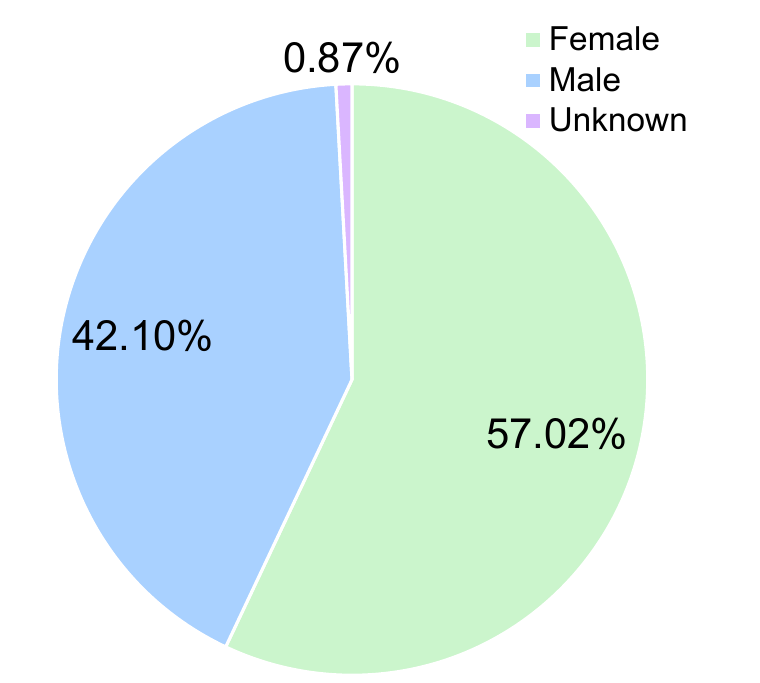}
    \caption{Gender distribution.}
    \label{fig:genderpie}
  \end{subfigure}\hfill
  \begin{subfigure}{0.5\linewidth}
    \centering
    \includegraphics[width=\linewidth]{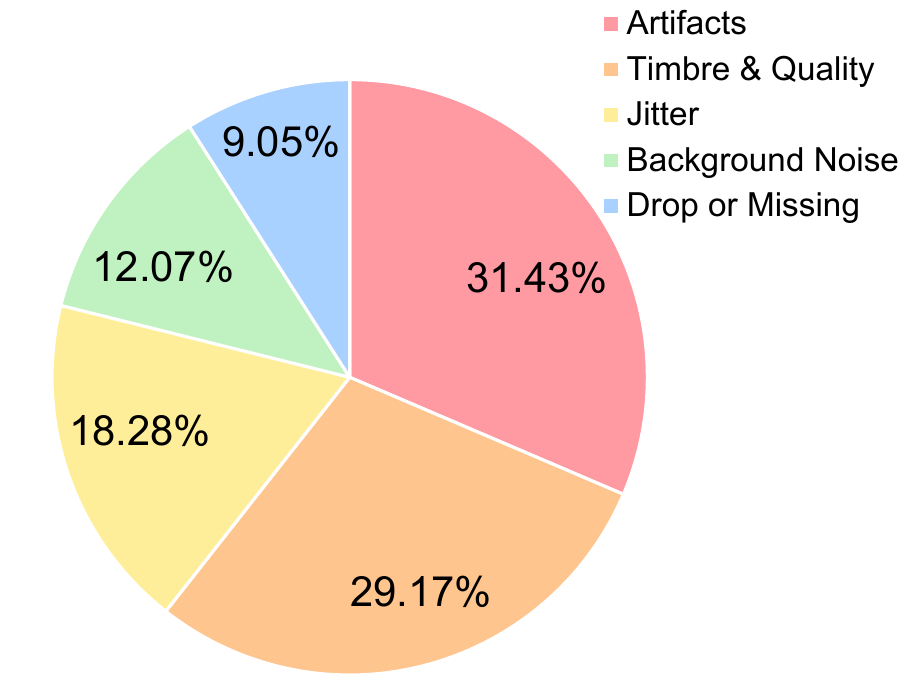} 
    \caption{Distortion type distribution.}
    \label{fig:distortionpie}
  \end{subfigure}

  \caption{Categorical metadata statistics in the SpeechEval assessment data: (a) Emotion distribution, with non-neutral emotions detailed on the right; (b) Gender distribution; (c) Distortion type distribution.}
  \label{fig:attribute_distribution}
\end{figure}

Figure~\ref{fig:score_distribution} presents the distribution of assessment scores across the eight annotated dimensions in the SpeechEval dataset. Most dimensions, including Overall Quality, Intelligibility, Distortion, and Subjective Experience, exhibit a reasonably balanced score distribution across the five-point scale, which reflects the diversity and coverage of our collected speech samples. Notably, the Speech Rate dimension shows a skewed distribution towards the extreme values, with a significant peak at score 5. This is due to the annotation scale for Speech Rate ranging from 1 (too slow) to 5 (too fast), where deviations from the optimal speed tend to be penalized more heavily. The score distribution indicates that our dataset provides rich and varied supervision signals, which are crucial for training and evaluating speech models across different perceptual aspects.

In Figure~\ref{fig:attribute_distribution}, we further analyze the categorical metadata statistics within the speech quality assessment task. The majority of samples are labeled as Neutral in terms of emotion, but non-neutral emotions are also represented, enhancing the emotional diversity of the dataset. Gender labels are balanced between male and female speakers, with a small portion marked as unknown. In terms of distortion types, the dataset includes a mix of artifacts, timbre issues, jitter, background noise, and missing segments, with artifacts being the most prevalent. These attribute distributions demonstrate that SpeechEval provides a comprehensive foundation for evaluating model robustness across realistic speech variations.

\subsection{Dataset Examples}

\begin{figure*}[htpb]
\begin{center}
\includegraphics[width=1\linewidth]{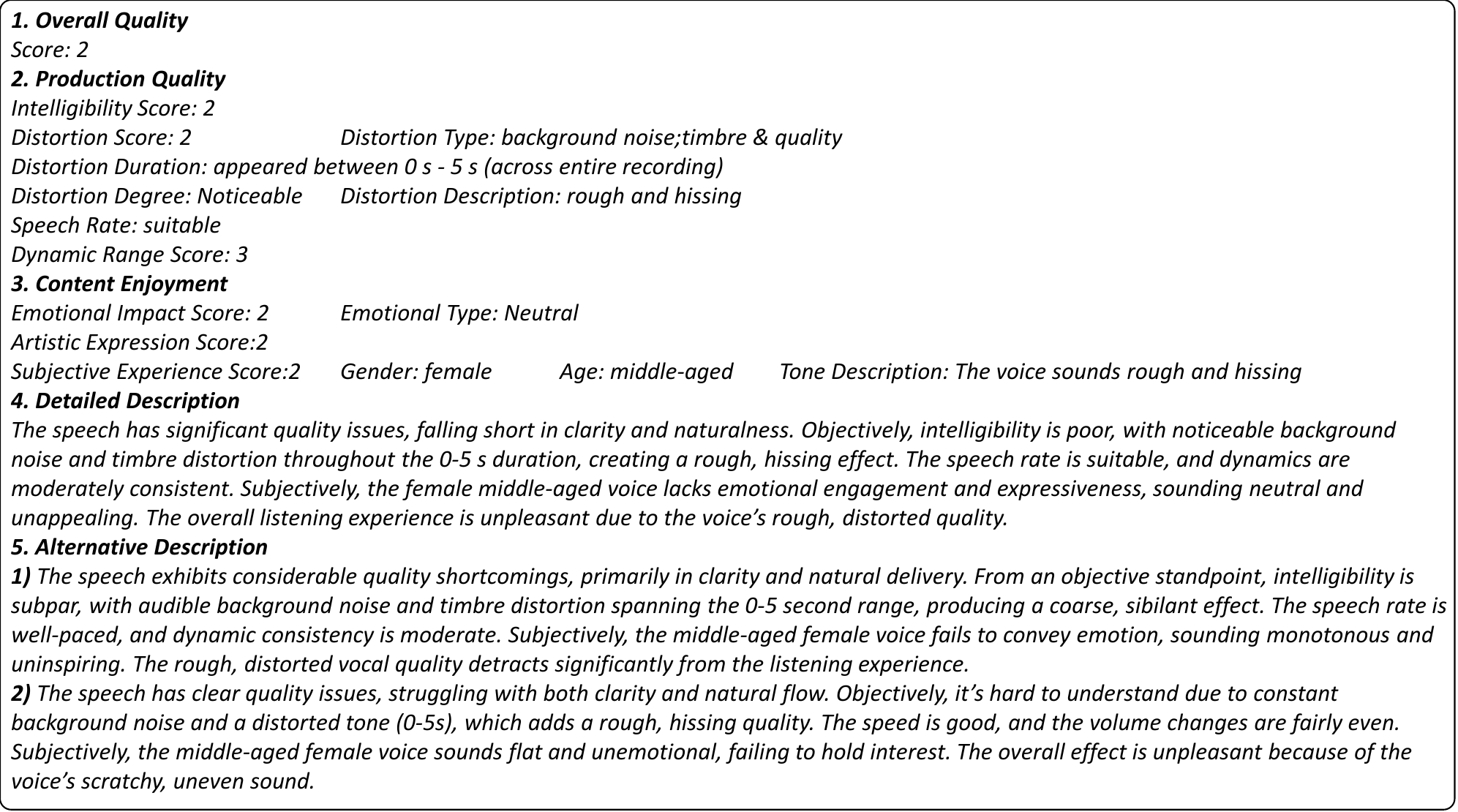}
\end{center}
\caption{Assessment example for a low-quality speech sample.}
\label{demo_Assessment_1}
\end{figure*}

\begin{figure*}[htpb]
\begin{center}
\includegraphics[width=1\linewidth]{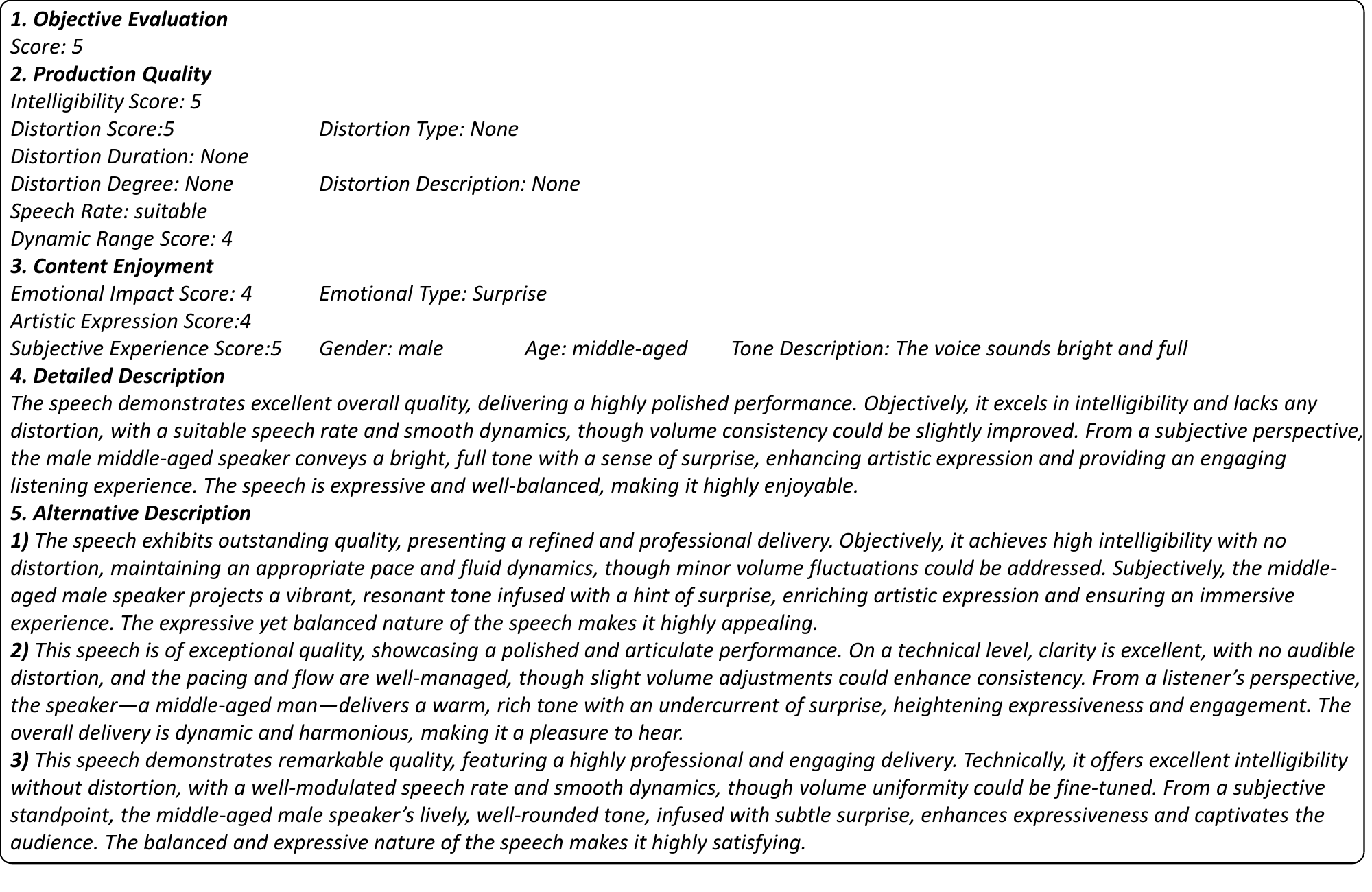}
\end{center}
\caption{Assessment example for a high-quality speech sample.}
\label{demo_Assessment_2}
\end{figure*}

\paragraph{Speech Quality Assessment}
A full evaluation for a low-quality speech sample is presented in Figure \ref{demo_Assessment_1}. Overall and intelligibility scores are low; continuous background noise and timbre/quality distortion are present across 0–5 s, described as rough and hissing. Rate and dynamics are acceptable, but emotional engagement is weak, and the tone sounds scratchy. The detailed and alternative descriptions explain how these distortions degrade clarity and listening experience.

As a counterpoint, Figure \ref{demo_Assessment_2} depicts a high-quality recording. Intelligibility is excellent, with a suitable speech rate and well-balanced dynamics; minor volume consistency could be improved. Subjectively, the voice is bright and full with a subtle “surprise” effect, yielding a highly pleasant listening experience.

\begin{figure*}[htpb]
\begin{center}
\includegraphics[width=1\linewidth]{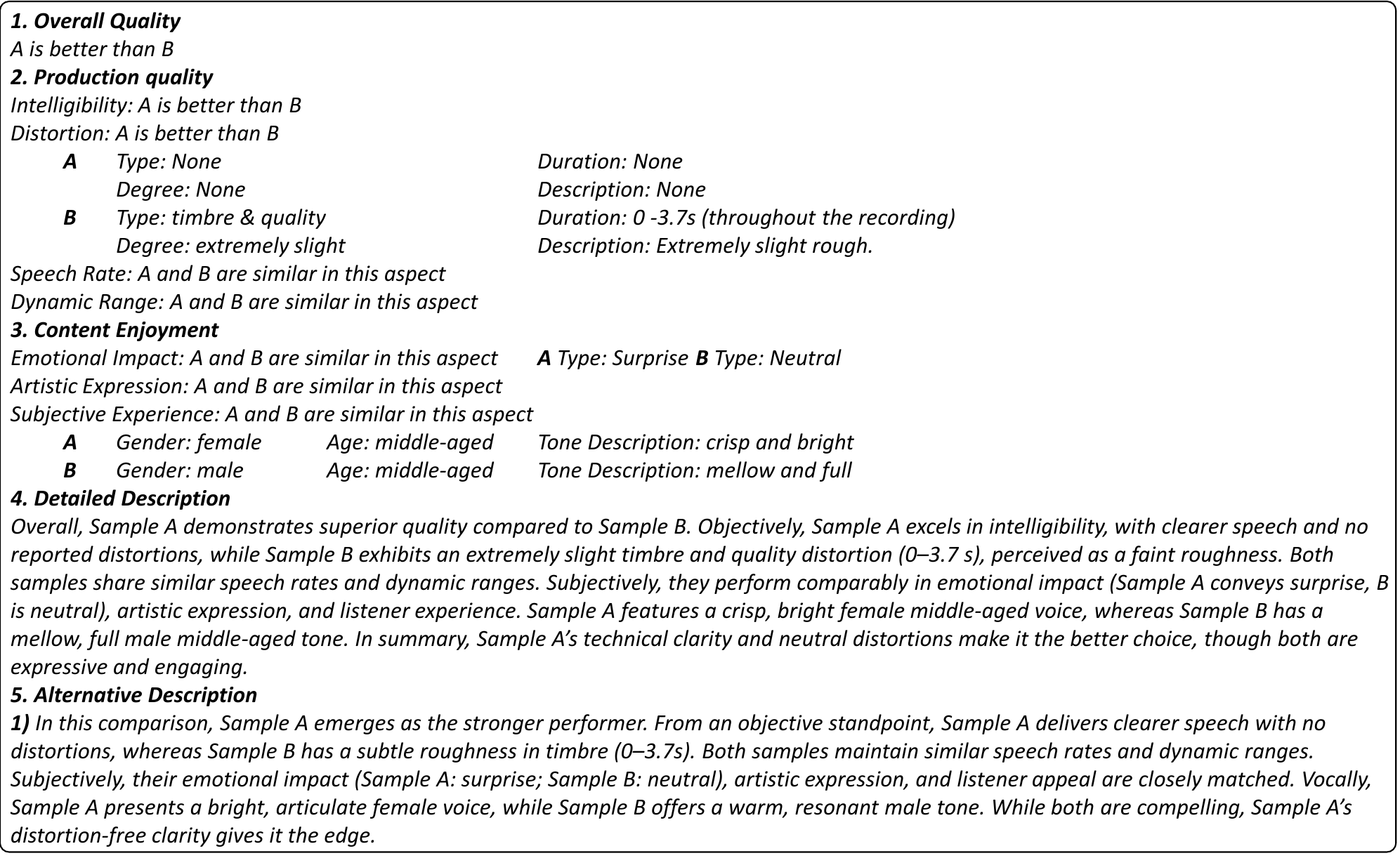}
\end{center}
\caption{Comparison example with a large quality gap between samples.}
\label{demo_Comparison_1}
\end{figure*}

\begin{figure*}[htpb]
\begin{center}
\includegraphics[width=1\linewidth]{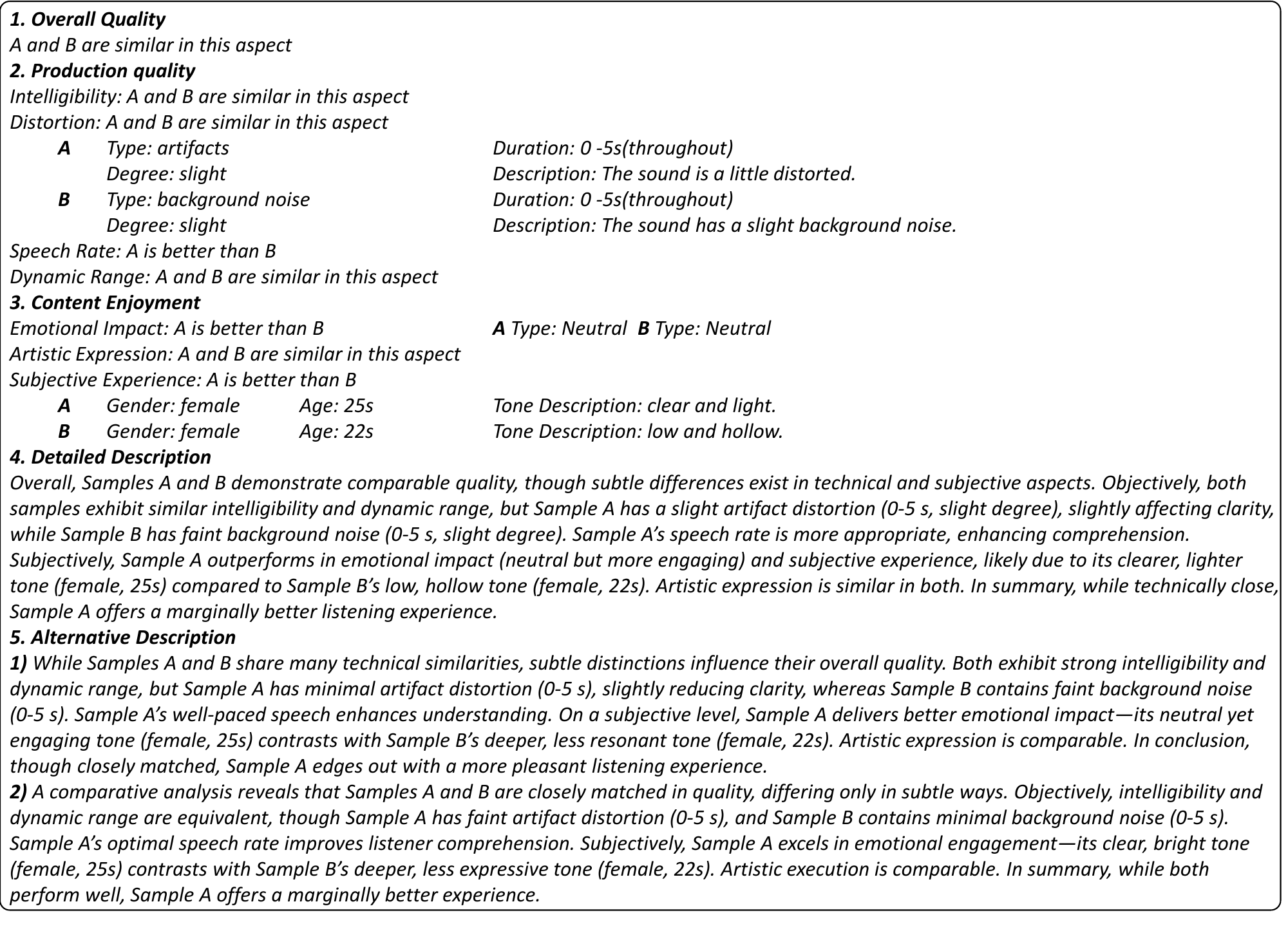}
\end{center}
\caption{Comparison example with a small quality gap between samples.}
\label{demo_Comparison_2}
\end{figure*}

\begin{figure*}[htpb]
\begin{center}
\includegraphics[width=1\linewidth]{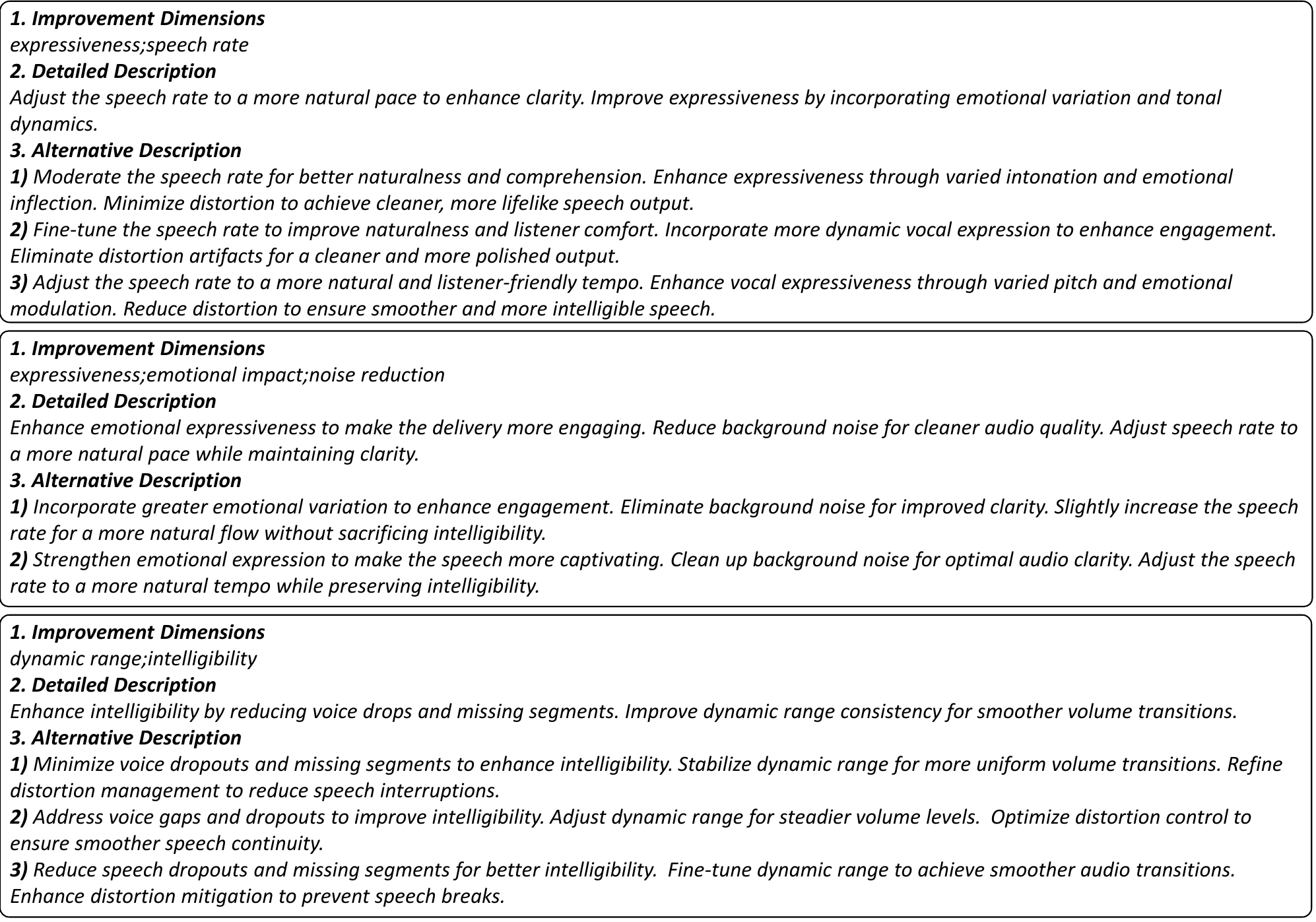}
\end{center}
\caption{Suggestion example for improving a low-quality speech sample.}
\label{demo_Suggestion_1}
\end{figure*}

\begin{figure*}[htpb]
\begin{center}
\includegraphics[width=1\linewidth]{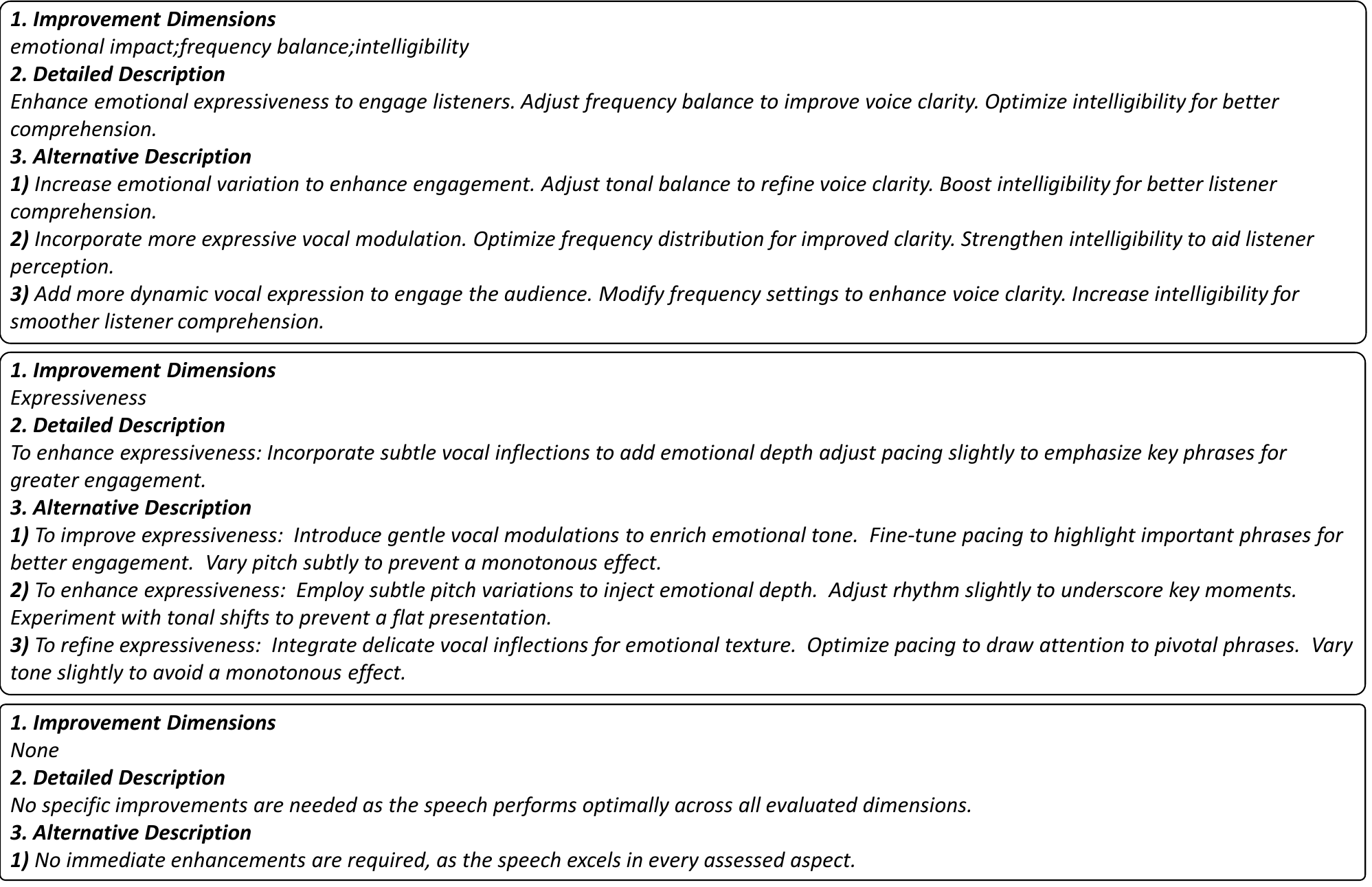}
\end{center}
\caption{Suggestion example for refining a high-quality speech sample.}
\label{demo_Suggestion_2}
\end{figure*}

\paragraph{Speech Quality Comparison}
In Figure~\ref{demo_Comparison_1}, the two samples differ markedly. Sample A is clearly superior in intelligibility and distortion control, while speech rate and dynamics are similar. Subjectively, A sounds crisp and bright with a livelier effect; B is mellow and resonant. The conclusion and alternatives emphasize that A’s technical clarity drives the overall preference.

Figure \ref{demo_Comparison_2} compares two recordings that are effectively on par across the core axes: intelligibility, dynamic range, and overall technical execution. Yet they diverge slightly across sub-dimensions: noise signature, timbre, and expressiveness. These micro-differences yield a modest preference for A, though both remain essentially comparable.

\paragraph{Speech Quality Improvement Suggestion}
For actionable remediation, Figure \ref{demo_Suggestion_1} provides targeted recommendations for a low-quality recording. The proposed improvements emphasize cleaning the signal, normalizing pacing, and strengthening expressiveness to raise intelligibility and listener appeal. The detailed and alternative descriptions translate these goals into concrete, implementation-ready guidance.

Figure \ref{demo_Suggestion_2} targets high-quality recordings where only subtle refinements are needed. It recommends light enhancements to emotional expressiveness, gentle tonal rebalancing, and fine-tuning of articulation, with alternative phrasings provided for direct implementation. The figure also includes a standardized “no improvement needed” case for samples that already perform optimally across all dimensions, ensuring consistent annotation when refinement is unnecessary.

\subsection{Detailed Annotation Protocol}
\label{app:annotation_protocol}

\paragraph{Annotation schema.}
\label{app:annotation protocol}
To support interpretability and task diversity, we design a unified annotation framework covering three aspects: overall rating, objective production quality, and subjective content enjoyment. These aspects are operationalized into eight subdimensions, including intelligibility, distortion, speech rate, dynamic range, tone balance, emotional impact, artistic expression, and subjective experience. Depending on the evaluation task, annotators provide either five-point scalar ratings or pairwise comparisons. The complete scoring criteria are reported in Tables~\ref{tab:overall_protocol}--\ref{tab:content_protocol}.

Beyond the primary ratings, we collect categorical metadata for distortion type, emotion type, and speaker gender. We additionally include five open-ended fields: distortion duration, distortion severity, perceptual description, speaker age, and speaking tone, as summarized in Table~\ref{tab:meta_protocol} and Table~\ref{tab:open-ended fields}.



\begin{table*}[ht]
\centering
\small
\setlength{\tabcolsep}{4pt} 
\resizebox{1\textwidth}{!}{%
\begin{tabular}{m{2.15cm} m{\dimexpr\textwidth-2.15cm-2\tabcolsep\relax}}
    \midrule
    \textbf{Sub-dimensions} & \textbf{Annotation Criteria} \\
    \midrule
    \centering Overall Quality &
        \begin{tabular}[t]{@{}l @{ }l @{\hspace{0.5em}}p{0.72\linewidth}@{}}
            \textbf{1 point} & \textbf{(Extremely Poor)} & Basic standard for lowest quality. \\
            \textbf{2 points} & \textbf{(Poor)} & Below average quality with obvious defects. \\
            \textbf{3 points} & \textbf{(Average)} & Moderate quality meeting basic listening needs. \\
            \textbf{4 points} & \textbf{(Good)} & High quality with minor imperfections. \\
            \textbf{5 points} & \textbf{(Excellent)} & Top-tier quality with no perceivable defects. \\
        \end{tabular} \\
    \midrule
\end{tabular}

}
  \vspace{-2pt}
  \caption{Annotation Protocol for Overall Quality.}
  \label{tab:overall_protocol}
\end{table*}



\begin{table*}[ht]
\centering
\small
\setlength{\tabcolsep}{4pt} 
\resizebox{1\textwidth}{!}{%
\begin{tabular}{m{2.15cm} m{\dimexpr\textwidth-2.15cm-2\tabcolsep\relax}}
    \midrule
    \textbf{Sub-dimensions} & \textbf{Annotation Criteria} \\
    \midrule

    \centering Intelligibility &
        \begin{tabular}[t]{@{}l @{ }l @{\hspace{0.5em}}p{0.72\linewidth}@{}}
            \textbf{1 point} & \textbf{(Extremely Poor)} & Pronunciation is vague. Most words or sentences are difficult to recognize or easily misheard. Listeners must concentrate very hard or replay multiple times to understand. \\[3pt]
            \textbf{2 points} & \textbf{(Poor)} & General idea is barely understandable with many ambiguities. Speech becomes difficult to hear even in slightly noisy environments. \\[3pt]
            \textbf{3 points} & \textbf{(Average)} & Most words and sentences are understandable. Only a few words or complex sentences require extra concentration. \\[3pt]
            \textbf{4 points} & \textbf{(Good)} & Speech is clear and easy to understand even in general background noise. Only occasional minor vagueness appears. \\[3pt]
            \textbf{5 points} & \textbf{(Excellent)} & Speech is extremely clear and almost all content is understood immediately. Pronunciation is accurate and expression is detailed. \\
        \end{tabular} \\
    \midrule

    \centering Distortion &
        \begin{tabular}[t]{@{}l @{ }l @{\hspace{0.5em}}p{0.72\linewidth}@{}}
            \textbf{1 point} & \textbf{(Extremely Poor)} & Obvious and continuous noise or heavy synthetic artifacts, which seriously affect understanding. \\[3pt]
            \textbf{2 points} & \textbf{(Poor)} & Audible background noise or buzzing is present, and frequent noise, splicing traces, or jitter can be heard. \\[3pt]
            \textbf{3 points} & \textbf{(Average)} & Minor noise or occasional synthetic traces, no serious listening interference. \\[3pt]
            \textbf{4 points} & \textbf{(Good)} & Very slight noise or synthetic feeling, detectable only in rare moments. \\[3pt]
            \textbf{5 points} & \textbf{(Excellent)} & The background is clean or synthetic traces are almost undetectable. The overall sound is natural and realistic. \\
        \end{tabular} \\
    \midrule

    \centering Speech Rate &
        \begin{tabular}[t]{@{}l @{ }l @{\hspace{5.7em}}p{0.72\linewidth}@{}}
            \textbf{Slow} & & Obviously lower than average speed. \\
            \textbf{Slightly Slow} & & Slightly lower than common speed. \\
            \textbf{Appropriate} & & Matches normal speed, easy for most people to understand. \\
            \textbf{Slightly Fast} & & Slightly higher than conventional speed. \\
            \textbf{Fast} & & Significantly higher than average spoken speed. \\
        \end{tabular} \\
    \midrule

    \centering Dynamic Range &
        \begin{tabular}[t]{@{}l @{ }l @{\hspace{0.5em}}p{0.72\linewidth}@{}}
            \textbf{1 point} & \textbf{(Extremely Poor)} & Sharp volume fluctuations, seriously affecting understanding. \\
            \textbf{2 points} & \textbf{(Poor)} & Frequent unreasonable volume changes. \\
            \textbf{3 points} & \textbf{(Average)} & Acceptable overall volume, but sudden changes in some paragraphs. \\
            \textbf{4 points} & \textbf{(Good)} & Stable volume most of the time, occasional minor fluctuations. \\
            \textbf{5 points} & \textbf{(Excellent)} & Stable and balanced volume. \\
        \end{tabular} \\
    \midrule
\end{tabular}
}
  \vspace{-2pt}
  \caption{Annotation Protocol for Production Quality.}
  \label{tab:quality_protocol}
\end{table*}

\begin{table*}[ht]
\centering
\small
\setlength{\tabcolsep}{4pt} 
\renewcommand{\arraystretch}{1.15} 
\resizebox{1\textwidth}{!}{%
\begin{tabular}{m{2.15cm} m{\dimexpr\textwidth-2.15cm-2\tabcolsep\relax}}
    \midrule
    \textbf{Sub-dimensions} & \textbf{Annotation Criteria} \\
    \midrule

    \centering \makecell{Emotional\\Impact} &
        \begin{tabular}[t]{@{}l @{ }l @{\hspace{0.5em}}p{0.72\linewidth}@{}}
            \textbf{1 point} & \textbf{(Extremely Poor)} & Strong machine-like feeling, no emotional fluctuation. \\
            \textbf{2 points} & \textbf{(Poor)} & Slight emotion, but still plain. \\
            \textbf{3 points} & \textbf{(Average)} & Adds emotional color to some paragraphs, but not sufficient. \\
            \textbf{4 points} & \textbf{(Good)} & Natural and smooth emotions, obvious emotional resonance. \\
            \textbf{5 points} & \textbf{(Excellent)} & Perfectly matches text atmosphere, deeply touches listeners with strong appeal. \\
        \end{tabular} \\
    \midrule

    \centering \makecell{Artistic\\Expression} &
        \begin{tabular}[t]{@{}l @{ }l @{\hspace{0.5em}}p{0.72\linewidth}@{}}
            \textbf{1 point} & \textbf{(Extremely Poor)} & Speech sounds mechanical and disconnected from the context. There is no sense of rhythm or layered meaning. \\[3pt]
            \textbf{2 points} & \textbf{(Poor)} & The speech shows only basic text expression and lacks contextual processing. \\
            \textbf{3 points} & \textbf{(Average)} & Limited rhythm/stress variation, beginning to express the text’s mood. \\
            \textbf{4 points} & \textbf{(Good)} & Integrates rhythm, context, and meaning well, with artistic appeal. \\
            \textbf{5 points} & \textbf{(Excellent)} & Rhythm, context, and nuance perfectly align, creating strong artistic tension. \\
        \end{tabular} \\
    \midrule

    \centering \makecell{Subjective\\Experience} &
        \begin{tabular}[t]{@{}l @{ }l @{\hspace{0.5em}}p{0.72\linewidth}@{}}
            \textbf{1 point} & \textbf{(Extremely Poor)} & Uncomfortable or boring, cannot be endured for a long time. \\
            \textbf{2 points} & \textbf{(Poor)} & Barely audible, lacks attractiveness and comfort. \\
            \textbf{3 points} & \textbf{(Average)} & Acceptable, neither boring nor pleasant. \\
            \textbf{4 points} & \textbf{(Good)} & Comfortable and natural, listeners are willing to continue. \\
            \textbf{5 points} & \textbf{(Excellent)} & Extremely pleasant, leaves a deep and happy impression. \\
        \end{tabular} \\
    \midrule
\end{tabular}
}
  \vspace{-2pt}
  \caption{Annotation Protocol for Content Enjoyment.}
  \label{tab:content_protocol}
\end{table*}

\begin{table*}[ht]
\centering
\small
\setlength{\tabcolsep}{4pt}
\renewcommand{\arraystretch}{1.15}
\resizebox{1\textwidth}{!}{%
\begin{tabular}{m{2.15cm} m{\dimexpr\textwidth-2.15cm-2\tabcolsep\relax}}
    \midrule
    \textbf{Sub-dimensions} & \textbf{Annotation Criteria} \\
    \midrule

    Distortion Type &
        \begin{tabular}[t]{@{}l @{\hspace{0.5em}}p{0.72\linewidth}@{}}
            \textbf{Background Noise} & Continuous or intermittent environmental noise. \\
            \textbf{Jitter} & Tremor or slight pitch flutter. \\
            \textbf{Drop / Missing} & Dropped frames or truncation, clipped words, brief silences or jumps. \\
            \textbf{Timbre \& Quality} & Unnatural tone. \\
            \textbf{Artifacts} & Non-natural sounds. \\
        \end{tabular} \\
    \midrule

    Emotion Type &
        \begin{tabular}[t]{@{}l @{\hspace{3.85em}}p{0.72\linewidth}@{}}
            \textbf{Happiness} & Higher pitch, stronger energy, slightly faster tempo, bright tone. \\
            \textbf{Sadness} & Lower pitch, softer loudness, slower rate, subdued affect. \\
            \textbf{Anger} & Strong energy, laryngeal tension, increased sharpness or bursts. \\
            \textbf{Fear} & More tremor, rising terminals, unstable delivery. \\
            \textbf{Disgust} & Nasal timbre, curt phrasing, flattened intonation. \\
            \textbf{Surprise} & Sudden increase in pitch and loudness, with clear and abrupt emphasis. \\
            \textbf{Neutral} & Pitch, energy, and rhythm remain stable, showing minimal emotional cues. \\
        \end{tabular} \\
    \midrule

    Speaker Gender &
        \begin{tabular}[t]{@{}l @{\hspace{4.15em}}p{0.72\linewidth}@{}}
            \textbf{Male} & Perceived masculine timbre. \\
            \textbf{Female} & Perceived feminine timbre. \\
            \textbf{Unknown} & Indeterminate. \\
        \end{tabular} \\
    \midrule
\end{tabular}
}
\vspace{-2pt}
\caption{Annotation Protocol for three types of categorical metadata.}
\label{tab:meta_protocol}

\end{table*}

\begin{table*}[ht]
\centering
\small
\setlength{\tabcolsep}{4pt}
\renewcommand{\arraystretch}{1.15}
\resizebox{1\textwidth}{!}{%
\begin{tabular}{m{2.15cm} m{\dimexpr\textwidth-2.15cm-2\tabcolsep\relax}}
    \midrule
    \textbf{Sub-dimensions} & \textbf{Annotation Criteria} \\
    \midrule

    \centering Distortion Severity &
        \begin{tabular}[t]{@{}l@{}}
        Explanation: The perceived intensity or impact level of the distortion.\\
        Example: \textit{"Slight."} / \textit{"Noticeable."} / \textit{"Severe."}
        \end{tabular} \\
    \midrule

    \centering\makecell{Perceptual\\Description} &
        \begin{tabular}[t]{@{}l@{}}
        Explanation: Short free-form notes describing the perceptual impression.\\
        Example: \textit{"The voice has poor timbre."}
        \end{tabular} \\
    \midrule

    \centering Speaker Age &
        \begin{tabular}[t]{@{}l@{}}
        Explanation: Estimated or perceived age range of the speaker.\\
        Example: \textit{"Thirties."}
        \end{tabular} \\
    \midrule

    \centering Speaking Tone &
        \begin{tabular}[t]{@{}l@{}}
        Explanation: Brief description of tone style.\\
        Example: \textit{"The voice sounds mellow but not clear enough."}
        \end{tabular} \\
    \midrule
\end{tabular}
}
\vspace{-2pt}
\caption{Annotation Protocol for five open-ended fields.}
\label{tab:open-ended fields}
\end{table*}

\paragraph{Annotators, compensation, and quality control.}
All annotators had relevant language backgrounds, with English proficiency at least at the College English Test Band 4 (CET-4) level (including TEM-8 holders and overseas students), and \(\geq\)1 year of annotation experience. A total of 56 professional annotators participated in this study, including 11 Chinese, 13 Japanese, 12 French, and 20 English annotators. The annotator pool comprised 40 female and 16 male annotators, with a relatively consistent age distribution across languages (mean age: 25.3--25.9 years; overall range: 23--31 years). For each target language, annotations were performed by native speakers or professionally trained annotators with domain expertise; among them, professionals in English, Japanese, and French accounted for 50\% of the workforce. 

All annotators completed training and passed a qualification test before participating in formal labeling. Annotators were fairly compensated: for Chinese, Japanese, and English speech, the payment was 0.62 USD per single-sample annotation and 0.83 USD per comparison task; for French speech, the rates were 0.69 USD and 0.90 USD, respectively. We employed an end-to-end platform-based pipeline with 100\% quality inspection; ambiguous cases were escalated to a third annotator for adjudication. We further integrated process-level quality-control tools, including lexicon checks for structured attributes (e.g., speaker gender and age) and consistency checks between distractor-item scores and selected distractor categories, to reduce annotation errors.

\section{Model}
\label{cotdata}

\subsection{Unified Prompt Format for SQ-LLM}

To support a unified evaluation across four distinct tasks, quality assessment, comparison, suggestion, and deepfake detection, we design standardized prompt templates to guide SQ-LLM’s behavior. These templates ensure consistent instruction formats while accommodating the unique requirements of each task. The following presents the representative prompts.

\vspace{2pt}
\textbf{Assessment Prompt}
\begin{lstlisting}[language=]
[Audio: <AUDIO_PLACEHOLDER>]
Please evaluate the overall quality of this speech.
\end{lstlisting}

\vspace{2pt}
\textbf{Comparison Prompt}
\begin{lstlisting}[language=]
Sample A: [Audio: <AUDIO_A_PLACEHOLDER>] 
Sample B: [Audio: <AUDIO_B_PLACEHOLDER>] 
Compare the speech quality of Sample A and Sample B. Please provide a comprehensive assessment.
\end{lstlisting}

\vspace{2pt}
\textbf{Suggestion Prompt}
\begin{lstlisting}[language=]
[Audio: <AUDIO_PLACEHOLDER>]
Please suggest specific aspects for improvement to enhance the overall quality of this speech.
\end{lstlisting}

\vspace{4pt}
\textbf{Deepfake Detection Prompt}
\begin{lstlisting}[language=]
[Audio: <AUDIO_PLACEHOLDER>]
Determine if this speech is real or a deepfake.
\end{lstlisting}

\subsection{Chain-of-Thought Example for Speech Quality Assessment}

To better illustrate the reasoning procedure , we provide an example of Chain-of-Thought  for the speech quality assessment task. In this case, the model is instructed to evaluate a single speech clip by assigning scores to eight perceptual dimensions, accompanied by brief justifications. The \texttt{<think>} block captures the internal reasoning process with structured judgments, while the \texttt{<answer>} block summarizes the evaluation in fluent natural language.

\vspace{8pt}
{\itshape 

<think>

For each dimension, give a score and a short explanation.

Overall Quality: 1/5

Intelligibility: 1/5

Distortion: 1/5 (timbre and quality;artifacts)

Speech Rate: suitable

Dynamic Range: 2/5

Emotional Impact: 1/5 (Neutral)

Artistic Expression: 1/5

Subjective Experience: 2/5 (male, middle-aged)

</think>

<answer>

The speech suffers from extremely poor overall quality. Objectively, intelligibility is severely compromised, with timbre and quality distortions lasting 0-3 seconds, making the vocal track unclear. The speech rate is suitable, but the dynamic range lacks smoothness. Subjectively, the neutral emotional impact and weak artistic expression fail to engage. The middle-aged male voice sounds hollow and thick, offering an unremarkable listening experience. In summary, the speech is marred by severe distortion, weak expressiveness, and a dull delivery.

</answer>}

\subsection{Chain-of-Thought Example for Speech Quality Comparison}

The following is an example of Chain-of-Thought reasoning used in the speech quality comparison task. The \texttt{<think>} section captures the intermediate thought process, where the model compares two speech clips dimension by dimension, offering judgments (e.g., ``A and B are similar'' or ``B is better than A''). This structured comparison allows the model to form a reasoned basis for its conclusion. The \texttt{<answer>} section then synthesizes these observations into a natural language summary that reflects the dimension-wise analysis.

\vspace{4pt}

{\itshape <think>

Compare the two audio samples across different quality dimensions.

Overall Quality: A and B are similar

Intelligibility: A and B are similar

Distortion: A and B are similar

Speech Rate: A and B are similar

Dynamic Range: A and B are similar

Emotional Impact: B is better than A

Artistic Expression: B is better than A

Subjective Experience: A and B are similar

</think>

<answer>

Overall, both speech samples (A and B) demonstrate comparable quality. Objectively, they are equally intelligible, with similar speech rates, dynamic range, and minimal distortion, though B has slightly better timbre. From a subjective perspective, B excels in emotional impact and artistic expression, delivering a more engaging performance, while the listening experience remains similar for both. In summary, B stands out slightly due to its expressive strengths, though both are technically balanced.

</answer>
}

\begin{figure*}[h]
\centering
\makebox[\textwidth][c]{%
  \includegraphics[width=\textwidth]{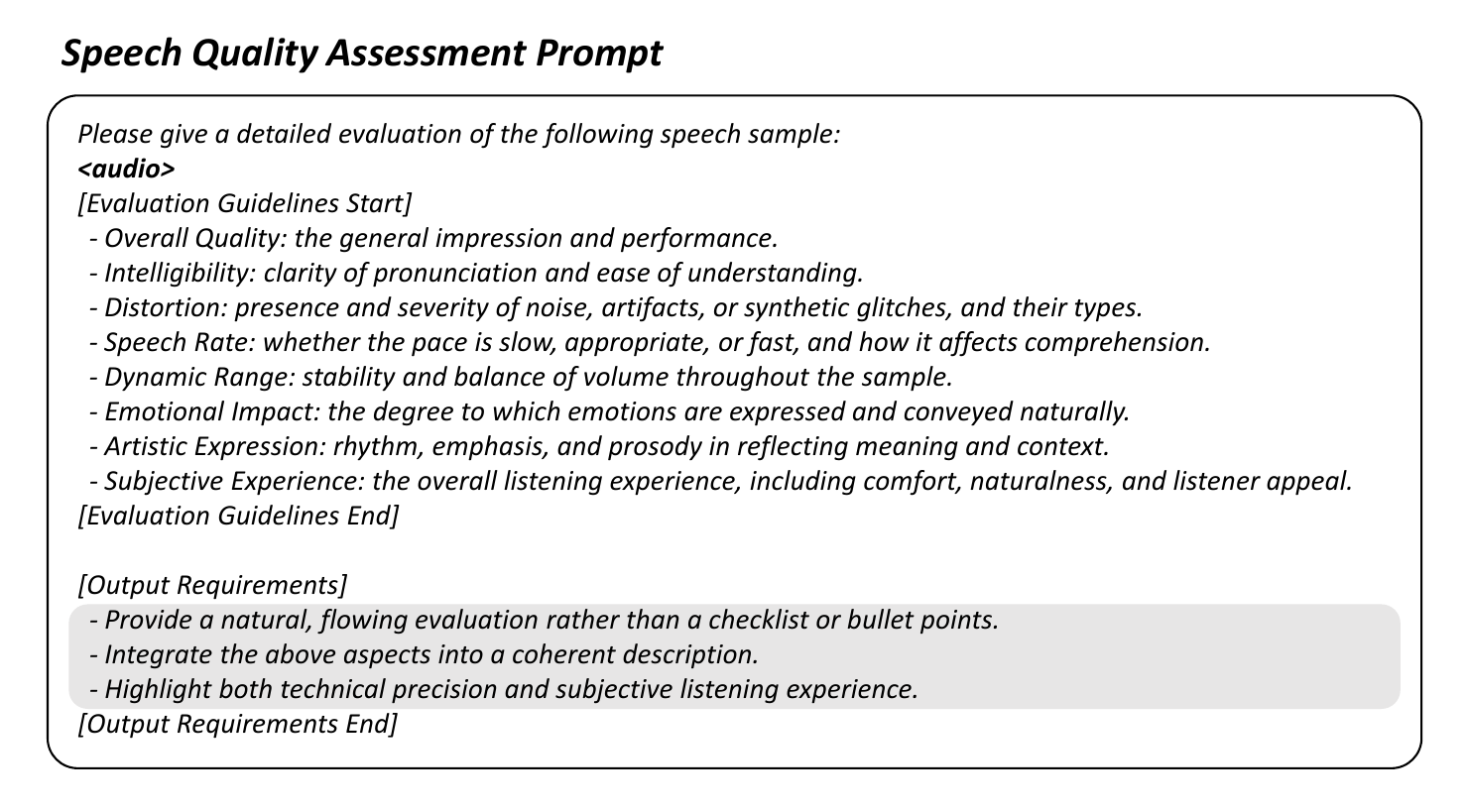}%
}
\caption{Speech Quality Assessment Prompt.}
\label{fig:sqa-prompt}
\end{figure*}

\begin{figure*}[h]
\centering
\makebox[\textwidth][c]{%
  \includegraphics[width=\textwidth]{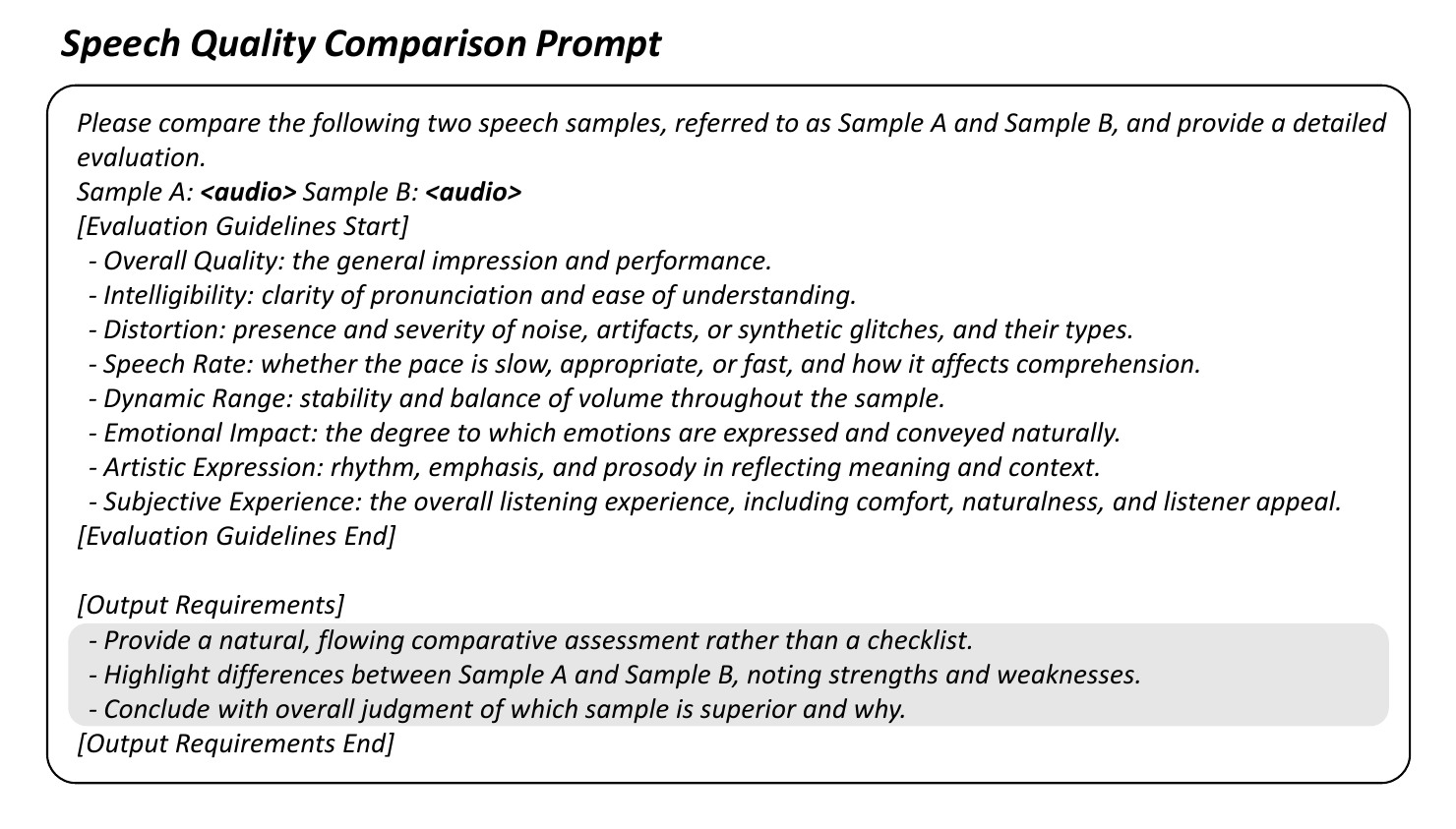}%
}
\caption{Speech Quality Comparison Prompt.}
\label{fig:sqc-prompt}
\end{figure*}

\begin{figure*}[h]
\centering
\makebox[\textwidth][c]{%
  \includegraphics[width=\textwidth]{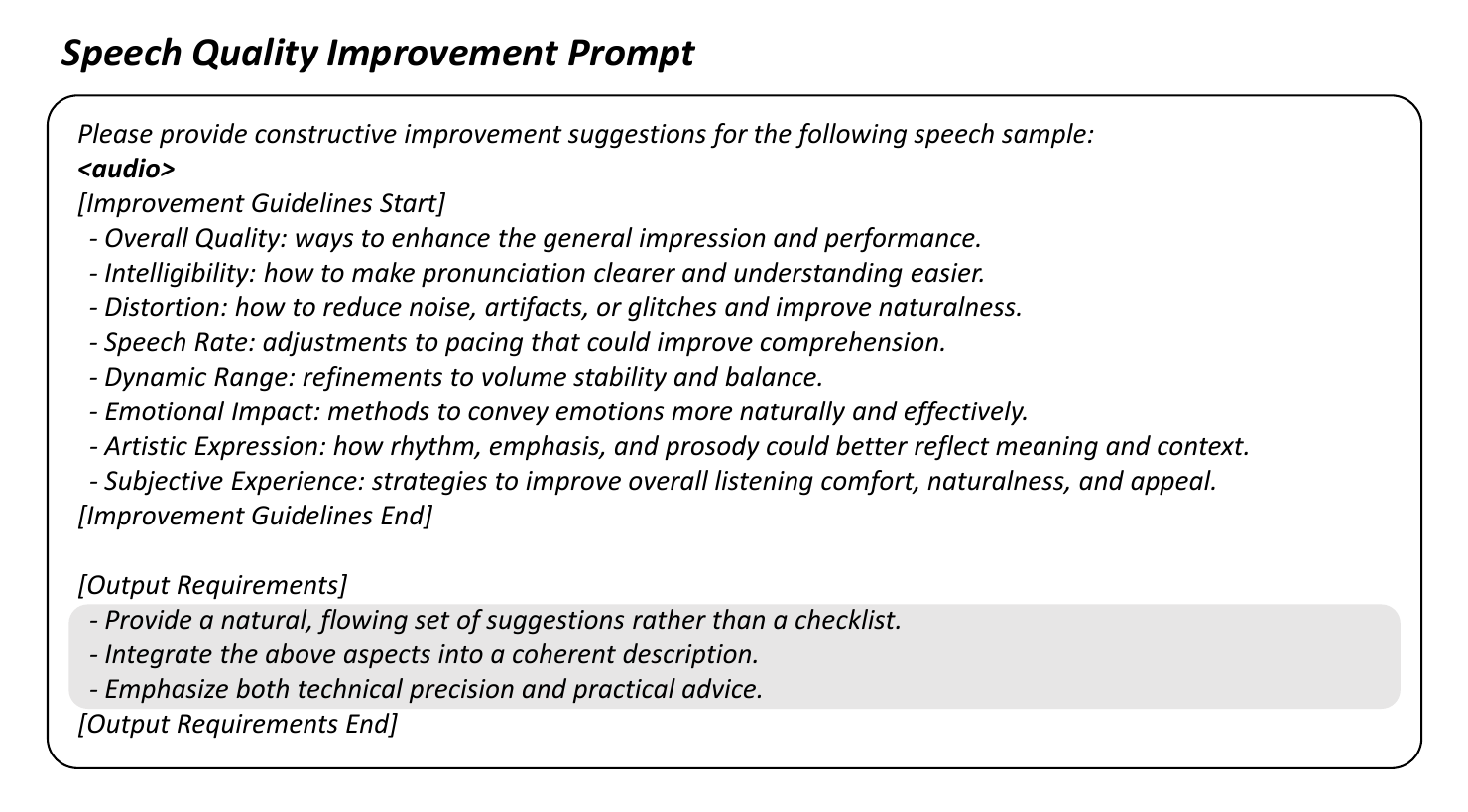}%
}
\caption{Speech Quality Improvement Prompt.}
\label{fig:sqi-prompt}
\end{figure*}

\begin{figure}[t]
\begin{center}
\includegraphics[width=1\linewidth]{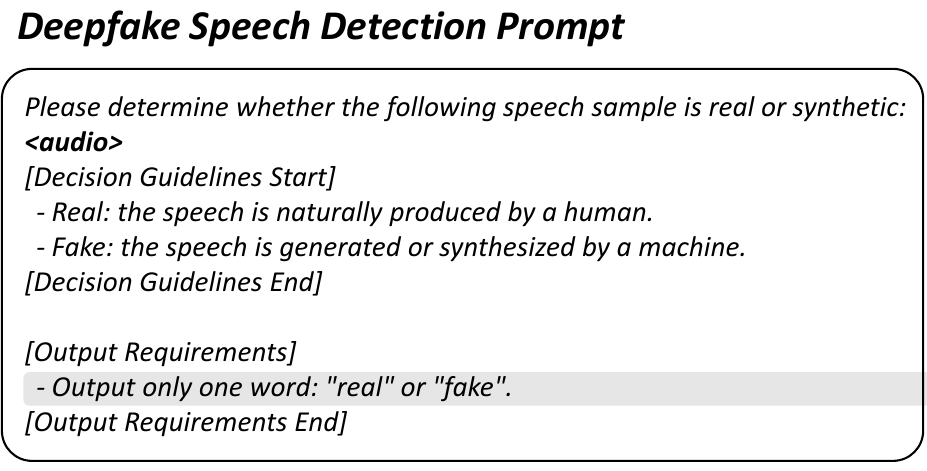}
\end{center}
\caption{Deepfake Speech Detection Prompt.}
\label{fig:dsd-prompt}
\end{figure}

\begin{figure*}[h]
\centering
\makebox[\textwidth][c]{%
  \includegraphics[width=1\textwidth]{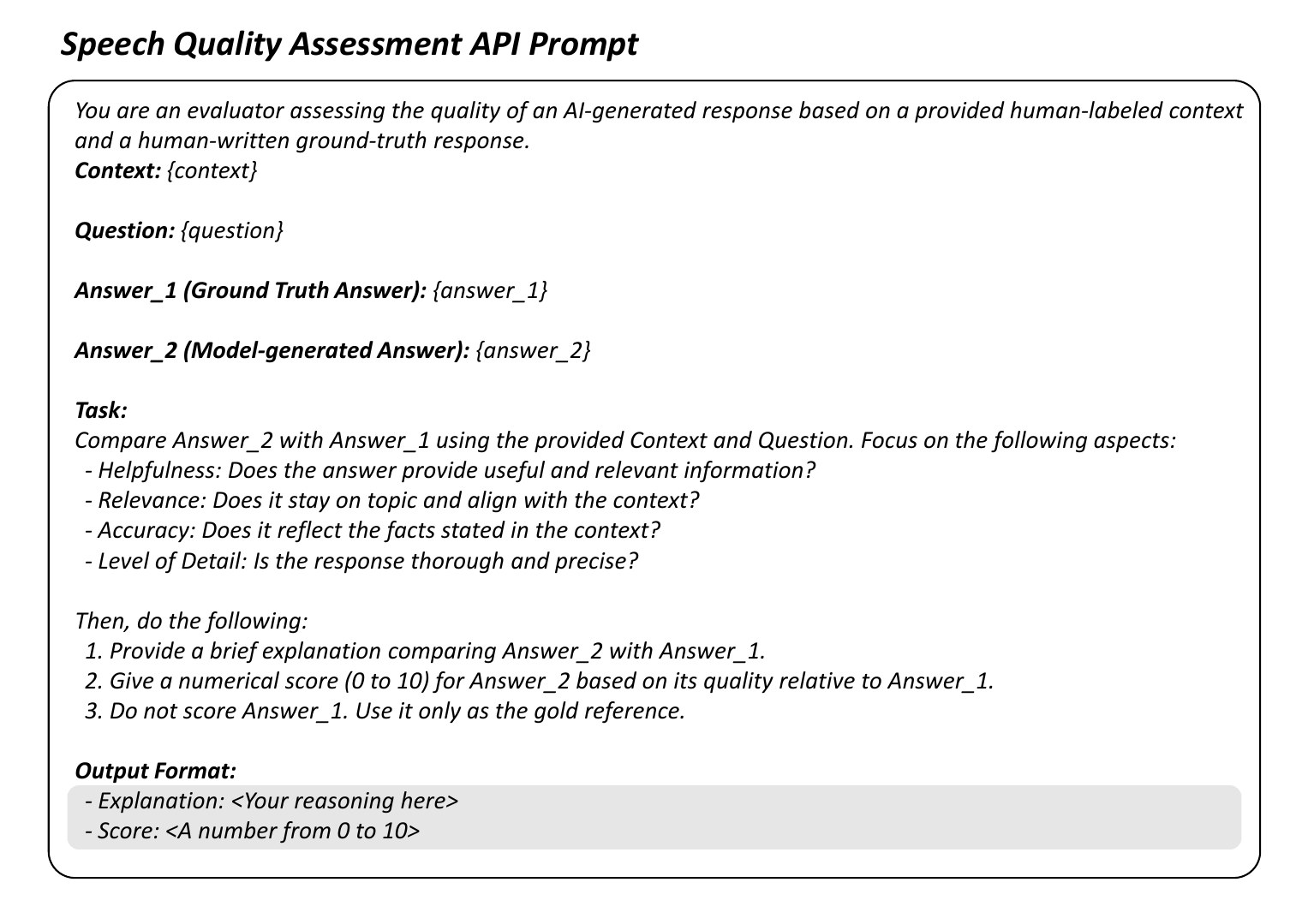}%
}
\caption{Speech Quality Assessment API Prompt.}
\label{fig:sqa-api-prompt}
\end{figure*}

\begin{figure}[t]
\begin{center}
\includegraphics[width=\linewidth]{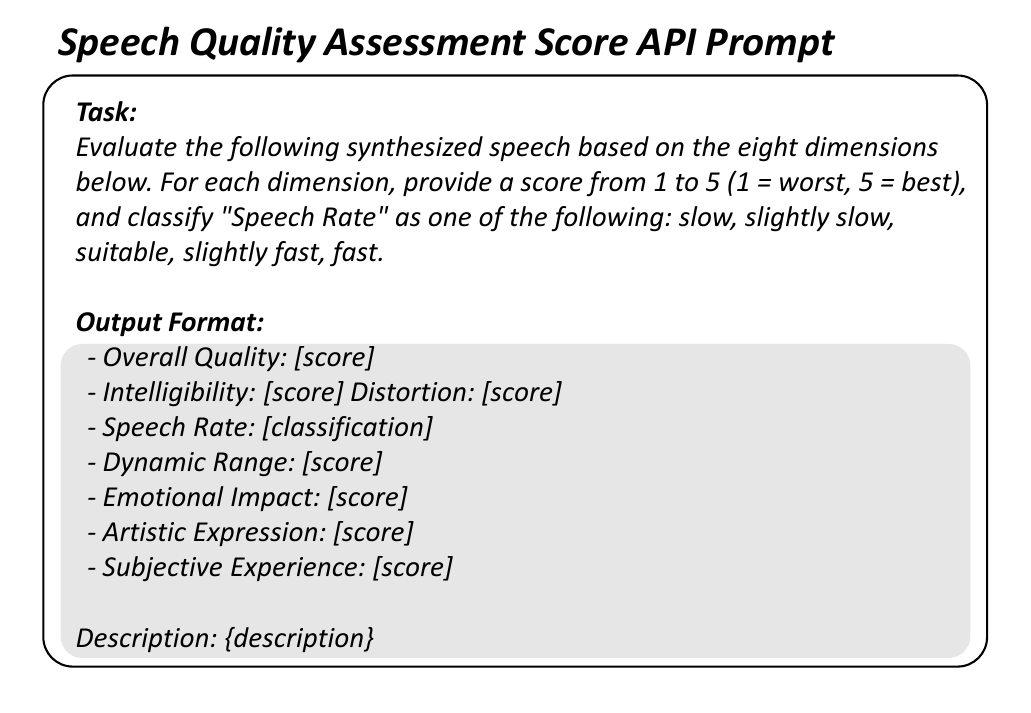}
\end{center}
\caption{Speech Quality Assessment Score API Prompt.}
\label{fig:sqa-score-api-prompt}
\end{figure}

\begin{figure*}[h]
\centering
\includegraphics[width=\textwidth]{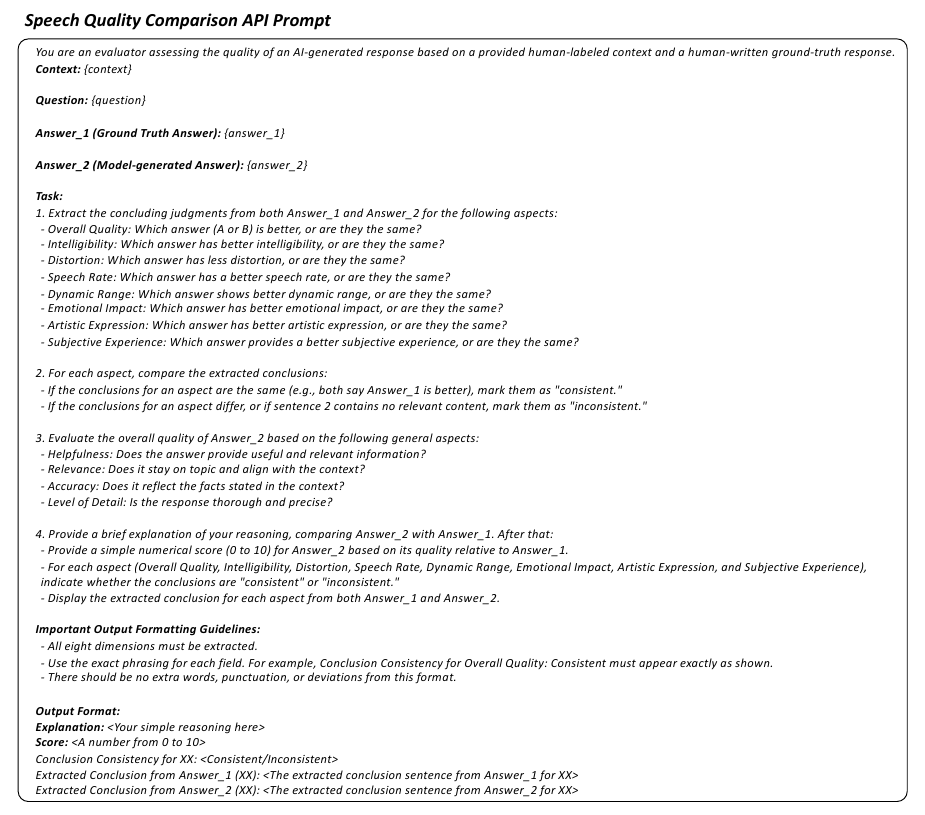}
\caption{Speech Quality Comparison API Prompt.}
\label{fig:sqc-api-prompt}
\end{figure*}

\subsection{Reward Criteria}

The four reward criteria are applied to three task formats (SingleEval, Suggest, CompareEval). Each evaluator outputs an integer in $[0,10]$. 

\textbf{[Helpfulness]}

\textbf{Assessment} \quad
\textbf{0–2} Barely helpful or misleading; advice doesn’t fit the context or could make results worse.
\textbf{2–4} Mostly generic advice (e.g., “improve clarity/intonation”) with little operational detail; not readily applicable. 
\textbf{4–6} Partly useful: identifies issues (distortion, limited emotional engagement) but lacks precise locations/parameters/practice details. 
\textbf{6–8} Actionable plan with minor gaps; mostly tailored; at least one check or fallback is suggested. 
\textbf{8–10} Directly usable and targeted: gives specific actions (e.g., reduce minor distortion at 0–0.5s, 1.3–2.1s, 3.1–3.8s via denoise/retake; adjust pacing; widen pitch/energy for expressiveness). 

\textbf{Comparison} \quad
\textbf{0–2} No decision, or misleading guidance that would pick the worse sample. 
\textbf{2–4} Vague preference (“A sounds nicer”) with minimal justification; weak for decision-making. 
\textbf{4–6} Gives a decision but reasons are partly generic or mixed; reader must infer the main takeaway. 
\textbf{6–8} Decision is usable with specific reasons, but lacks either prioritization, trade-offs, or implications. 
\textbf{8–10} Clear decision (A/B or justified tie) with ranked reasons that map to the brief’s goals; highlights where and why one sample is better (e.g., “A cleaner in 0–3.2 s; more balanced dynamics in verse lines”), plus what that implies (intelligibility, listening comfort). If close, states trade-offs and when to prefer each. 

\textbf{Suggestion} \quad
\textbf{0–2} Misleading or counterproductive suggestions; could worsen quality.
\textbf{2–4} Mostly platitudes (“be more expressive”); little practical value without extra work.
\textbf{4–6} Partly helpful: directions are generic or incomplete; user must figure out key details to execute.
\textbf{6–8} Solid and usable, but misses one element (priority, verification, or justification).
\textbf{8–10} Directly improves the piece: prioritized, actionable steps with clear how-to. 

\textbf{[Relevance]} 

\textbf{Assessment} \quad
\textbf{0–2} Largely off-topic or at odds with instructions/context. 
\textbf{2–4} Weak focus: substantial digressions or misses a key element from the context (e.g., ignores the provided distortion intervals). 
\textbf{4–6} Mostly relevant but includes nontrivial extraneous content or overlooks a minor stated point (e.g., underemphasizes dynamic range being stable). 
\textbf{6–8} Generally focused with brief, still-useful tangents; constraints are respected with only trivial deviations. 
\textbf{8–10} Stays tightly on clarity/intelligibility, pacing, distortion with timestamps, dynamic range stability, tone/timbre, expressiveness/emotion; no irrelevant digressions; follows the reference. 

\textbf{Comparison} \quad
\textbf{0–2} Off-topic or contradicts the comparison instruction (e.g., evaluates only A). 
\textbf{2–4} Weak alignment: talks about single-sample qualities without comparing, or ignores key axes. 
\textbf{4–6} Some relevant content, but includes nontrivial digressions or underplays a required axis. 
\textbf{6–8} Mostly on-topic with brief tangents that still inform the comparison. 
\textbf{8–10} Tightly focused on A vs B across the specified axes (clarity/intelligibility, pacing, distortion with timestamps, dynamic range, tone/expressiveness); no unrelated commentary; follows output constraints (e.g., compare rather than re-summarize background). 

\textbf{Suggestion} \quad
\textbf{0–2} Largely off-topic or contradicts instructions. 
\textbf{2–4} Weak alignment; misses key issues mentioned/implied in context. 
\textbf{4–6} Some relevance, but includes nontrivial digressions or overlooks a stated focus. 
\textbf{6–8} Mostly on-topic, with brief tangents that still help. 
\textbf{8–10} Tightly aligned to the observed issues (e.g., rate, emotional nuance, expressiveness) and any provided constraints; no off-topic or format violations (stays in “suggest” mode, not scoring/explaining). 

\textbf{[Accuracy]} 

\textbf{Assessment} \quad
\textbf{0–2} Predominantly incorrect or fabricated; contradicts core facts; unsafe/misleading. 
\textbf{2–4} Key error or contradiction; parts may remain usable. 
\textbf{4–6} Some incorrect details or unsupported claims, but the main idea is salvageable. 
\textbf{6–8} Minor imprecision that doesn’t change the outcome (e.g., “slight” vs. “minor”); overall alignment remains correct. 
\textbf{8–10} Fully consistent with the reference: correctly notes every details; no fabricated metrics or contradictions. 

\textbf{Comparison} \quad
\textbf{0–2} Largely inaccurate/fabricated; contradicts core facts or uses faulty audio concepts. 
\textbf{2–4} Major misread of evidence leading to doubtful verdict. 
\textbf{4–6} Some incorrect or unsubstantiated claims, but the main verdict is still plausible. 
\textbf{6–8} Minor imprecision that doesn’t change the outcome; overall alignment intact. 
\textbf{8–10} Consistent with context: correctly attributes issues. No fabricated timestamps/metrics; no internal contradictions; terminology used correctly. 

\textbf{Suggestion} \quad
\textbf{0–2} Fabricated issues, unsafe/incorrect advice, or contradictions with context. 
\textbf{2–4} Major mismatch or questionable technique likely to underperform. 
\textbf{4–6} Some mismatches yet partially useful. 
\textbf{6–8} Minor imprecision but generally appropriate; outcomes remain valid.
\textbf{8–10} Suggestions match the evidence (no invented problems), choose appropriate, safe remedies, and avoid technical/myth errors. 

\textbf{[Level of Detail]} 

\textbf{Assessment} \quad
\textbf{0–2} Little to no actionable detail; mostly restates the prompt or uses vague descriptors. 
\textbf{2–4} High-level outline; major gaps. 
\textbf{4–6} Main points present but lacks important parameters; some ambiguity remains. 
\textbf{6–8} Strong detail with minor omissions (eg. missing one timestamp or a small validation step); still largely reproducible. 
\textbf{8–10} Reproducible detail: cites exact timestamps for issues; separates objective vs. subjective aspects. 

\textbf{Comparison} \quad
\textbf{0–2} Vague restatements; no actionable or verifiable detail. 
\textbf{2–4} High-level bullets with little localization or evidence; hard to verify. 
\textbf{4–6} Names main areas but have some ambiguity. 
\textbf{6–8} Specific and useful but missing one element. 
\textbf{8–10} Concrete evidence: cites timestamps/segments for differences; separates objective from subjective. 

\textbf{Suggestion} \quad
\textbf{0–2} Vague or filler; no operational detail.
\textbf{2–4} High-level bullets only; little to reproduce in practice. 
\textbf{4–6} Main areas named, but lacks crucial parameters/examples; ambiguity remains. 
\textbf{6–8} Detailed but missing one element. 
\textbf{8–10} Specific targets, concrete how-to steps and actionables.

\section{Experimental Setup}

\subsection{Data splitting details}
\label{app:data_split}
In the Deepfake Speech Detection task, the partitioning method of the dataset is crucial for effective model training and evaluation. The ASVspoof challenge was the first to establish the common practice of dataset construction for this task, which has since been widely adopted in subsequent research. Following this established practice, we partition the dataset for the Deepfake Speech Detection task accordingly. The dataset comprises a total of 32,207 samples, divided into training, validation, and test sets in proportions of 20.5\%, 17.8\%, and 61.7\%, respectively. To better simulate real-world applications, the proportion of real samples gradually decreases across the subsets, with rates of 33.1\%, 22.7\%, and 20.3\%, ensuring the adequacy and diversity of the training set and aligning with the challenging Zero-shot evaluation scenario, which is a key focus of this task. The specific partitioning strategies are as follows: for certain fake sample sources, a 5:5:0 train-validation split is used to assess the model’s ability to fit known spoofing distributions, while a portion of fake sample sources is fully assigned to the test set with a 0:0:10 ratio, evaluating the model's zero-shot generalization capability in the context of previously unseen spoofing scenarios.
Furthermore, due to significant variations in language and data scale across sample sources, an imbalance in training data arises. To mitigate this, we adopt four different partitioning ratios (2:2:6, 4:2:4, 6:2:2, and 1:1:8) for certain sample sources to ensure sufficient training samples while maintaining balanced distributions across the subsets. Finally, we verify that the resulting partitions are mutually exclusive across tasks, ensuring that no data overlap or leakage occurs.

\subsection{LLM-based Metrics}
\label{app:llmscore}

To evaluate the outputs produced by the three baseline models in the ms-swift framework, we employ the DeepSeek to score model inferences on the Speech Quality Assessment and Speech Quality Comparison prompt tasks, and quantify synthesized speech along multiple dimensions.

For the SQA task, we first assess the overall accuracy and quality of the generated paragraph-level descriptions. Concretely, human-annotated speech evaluation texts are used as ground-truth and, together with each baseline model’s generated output, are submitted to the DeepSeek API. We require DeepSeek to perform a comparative evaluation along the dimensions of Helpfulness, Relevance, Accuracy, and Level of Detail, returning a 10-point score and a textual justification for each baseline output. The prompt used for the paragraph-level assessment is given in Figure~\ref{fig:sqa-api-prompt}.

Next, we extract fine-grained dimensions from the paragraph-level descriptions and perform an accuracy assessment per dimension. As shown in Figure~\ref{fig:sqa-score-api-prompt}, each baseline model’s natural-language SQA output is sent to DeepSeek to be quantitatively mapped onto eight quality dimensions: Overall Quality, Intelligibility, Distortion, Speech Rate, Dynamic Range, Emotional Impact, Artistic Expression, and Subjective Experience. Speech Rate is treated as a categorical label with five levels: Slow, Slightly Slow, Appropriate, Slightly Fast, and Fast, while the other seven dimensions are mapped to a five-point numeric scale. The resulting quantitative scores are then compared against human ground-truth annotations to compute accuracy statistics.

For the SQC task, we follow an analogous procedure: DeepSeek is used to evaluate model outputs both at the paragraph level and across the same eight fine-grained dimensions, producing comparable numeric scores and textual explanations. The SQC prompt used to elicit these comparative judgments is shown in Figure~\ref{fig:sqc-api-prompt}, and the scores thus obtained are evaluated against ground-truth annotations to enable direct comparison across models.

\subsection{Deepfake Detection Metrics}
\label{app:deepfakemetric}

The ACC metric applies to all LLM outputs, regardless of whether the token at the expected position is semantically valid. A prediction is counted as correct only when the generated token matches the ground-truth response among the key responses (“Fake” and “Real”).
In contrast, the EER and minDCF are evaluated only on semantically valid outputs. Since LLMs do not provide explicit confidence scores, we approximate the posterior probabilities by applying a two-dimensional softmax over the logits of the key response tokens, following the method described in \citet{gu2025allm4add}.
Note that for expert systems, the EER and minDCF are computed directly from the inference scores, and the ACC is subsequently obtained by binarizing these scores into categorical labels.

\begin{table*}[t]

\centering
\resizebox{0.9\linewidth}{!}{%
{%
\renewcommand{\arraystretch}{1.2}%
\begin{tabular}{lcccccccc|c}
\hline
\multicolumn{1}{c}{\bf MODEL} &
\multicolumn{1}{c}{\bf OVR} &
\multicolumn{1}{c}{\bf INT} &
\multicolumn{1}{c}{\bf DST} &
\multicolumn{1}{c}{\bf DYN} &
\multicolumn{1}{c}{\bf EMO} &
\multicolumn{1}{c}{\bf EXP} &
\multicolumn{1}{c}{\bf SUBJ} &
\multicolumn{1}{c|}{\bf AVG PCC} &
\multicolumn{1}{c}{\bf SR ACC} \\
\hline
Qwen2-Audio-7B                      & 0.073 & 0.037 & 0.080 & 0.043 & 0.070 & 0.078 & 0.072 & 0.065 & 0.664 \\
Qwen2.5-Omni-7B                     & 0.070 & 0.078 & 0.070 & 0.019 & 0.057 & 0.050 & 0.106 & 0.064 & \bf{0.736} \\
MiDashengLM-7B                      & 0.158 & 0.114 & 0.138 & 0.071 & 0.156 & 0.102 & 0.159 & 0.128 & 0.712 \\
\hdashline
Qwen3-8B + Whisper                  & 0.424 & 0.387 & 0.470 & 0.274 & 0.363 & 0.329 & 0.418 & 0.381 & 0.676 \\
Qwen2.5 + Audiobox                  & 0.457 & 0.423 & 0.585 & 0.209 & 0.376 & 0.348 & 0.420 & 0.403 & 0.644 \\
Qwen3-4B + WavLM                    & 0.422 & 0.387 & 0.460 & 0.207 & 0.314 & 0.334 & 0.378 & 0.357 & 0.665 \\
FT Qwen2-Audio-7B            & 0.244 & 0.192 & 0.455 & 0.202 & 0.209 & 0.295 & 0.269 & 0.267 & 0.706 \\
\hdashline
\textbf{SQ-LLM (OURS)}                                & \bf{0.520} & \bf{0.505} & \bf{0.592} & \bf{0.329} & \bf{0.434} & \bf{0.378} & \bf{0.456} & \bf{0.459} & 0.731 \\
\hline
\end{tabular}%
}%

}
\caption{Detailed speech quality assessment across models: each dimension Pearson correlation coefficients (PCC) with human ratings and speech rate accuracy; abbreviations are OVR (overall quality), INT (intelligibility), DST (distortion), DYN (dynamics), EMO (emotional), EXP (expression), SUBJ (subjective). AVG PCC denotes the mean PCC over all dimensions, and SR ACC denotes speech rate accuracy.}
\label{tab:singleeval_pcc}
\end{table*}

\subsection{Baselines}
\label{app:baselines}

We evaluate four speech-related tasks, Speech Quality Assessment, Speech Quality Comparison, Speech Quality Improvement Suggestion, and Deepfake Speech Detection by using three models from the ms-swift framework: Qwen2-Audio-7B-Instruct~\citep{chu2024qwen20audio}, Qwen2.5-Omni-7B~\citep{xu2025qwen2050omni}, and MiDashengLM-7B~\citep{dinkel2025midashenglm0}.

Qwen2-Audio-7B-Instruct is a large-scale audio-language model that accepts various audio inputs and is optimized for instruction-following in both voice chat and audio analysis modes. Its audio encoder is based on the Whisper-large-v3 model.

Qwen2.5-Omni-7B is an end-to-end omni-multimodal model designed to perceive text, images, audio, and video. It uses a Thinker-Talker architecture to generate both text and streaming speech responses.

MiDashengLM-7B is an open-source model designed for efficient and comprehensive audio understanding by training on "general audio captions" that fuse speech, sound, and music information into a single textual representation.

These models have comparable parameter scales and all support mixed multimodal inputs consisting of multiple speech segments and text. During inference, we configure the models to generate textual outputs. The hyperparameters are set uniformly with max\_new\_tokens = 2048 and max\_batch\_size = 64, while all other parameters remain at their default settings.

For the Deepfake Speech Detection task, three baseline systems, RawNet2 \citep{tak2021end}, AASIST \citep{jung2022aasist}, and AASIST2 \citep{zhang2024improving}, are employed.  
RawNet2 and AASIST follow the official ASVspoof 5 Challenge implementations\footnote{https://github.com/asvspoof-challenge/asvspoof5}, while AASIST2 is re-implemented using the pretrained feature extractor \textit{voidful/wav2vec2-xlsr-multilingual-56} from Hugging Face. All systems retain their original architectures, and the input features and training setups are standardized to ensure a fair comparison.

RawNet2 uses a learnable Sinc front-end \citep{ravanelli2018speaker} with residual stacks for end-to-end learning on raw waveforms, incorporating filter-wise feature map scaling to enhance discriminability. It has been widely adopted for deepfake speech detection and serves as an official baseline in the ASVspoof 2021 challenge.  
AASIST augments a raw-waveform encoder with a spectro-temporal heterogeneous graph attention mechanism, coupled with max graph operations and an extended readout scheme to model artifacts introduced by synthesis and conversion.  
AASIST2 integrates a wav2vec2 pretrained feature extractor \citep{babu2021xls} with Res2Net-based multi-scale blocks \citep{gao2019res2net}, and applies dynamic chunking with adaptive large-margin fine-tuning to improve robustness across durations, particularly for short utterances.

For reproducibility, all models are trained using the Adam optimizer (learning rate = 0.0001, batch size = 24) with a class-weighted cross-entropy loss (negative $\mathbin{:}$ positive = $7 \mathbin{:} 3$). Training is conducted for up to 50 epochs on a single NVIDIA RTX 4090 GPU, with early stopping based on validation EER and a patience of 5.

\begin{figure*}
    \centering
    \includegraphics[width=0.8\linewidth]{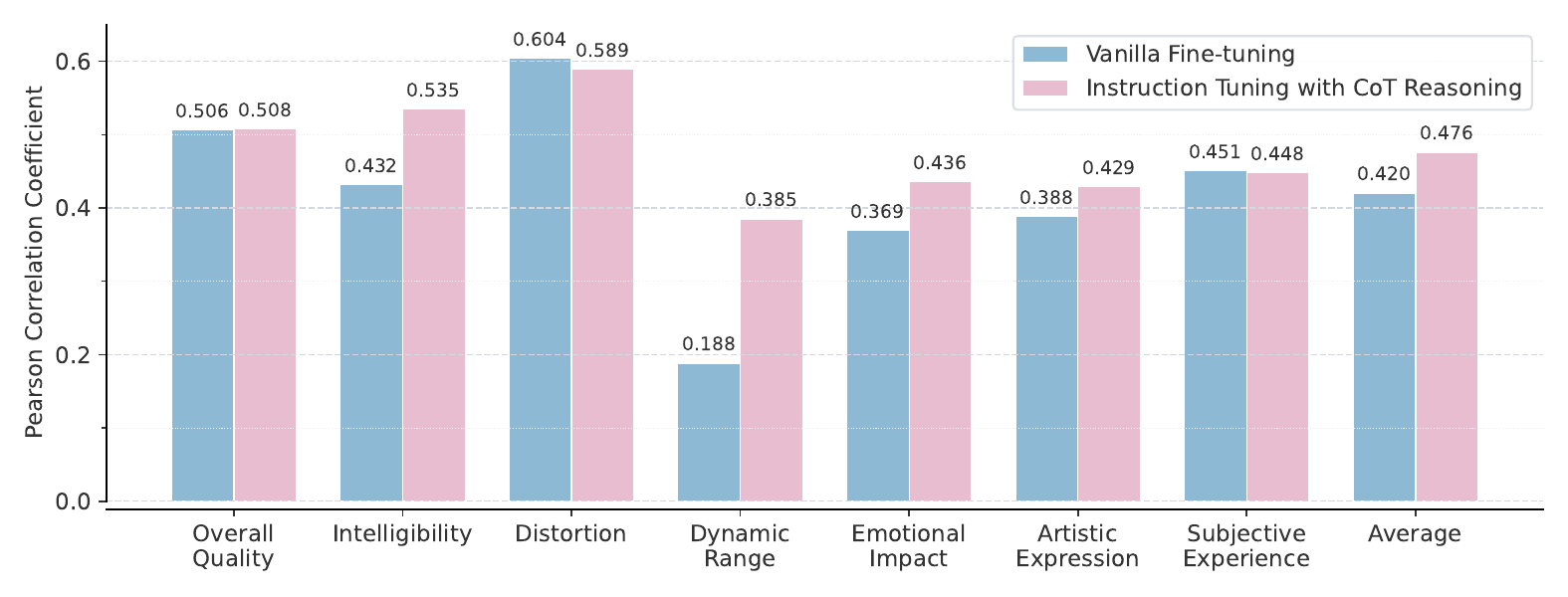}
    \caption{Pearson correlation coefficients of SQ-LLM predictions with human ratings across dimensions. The comparison is made between vanilla fine-tuning and instruction tuning with Chain-of-Thought reasoning.}
    \label{fig:cot_pcc}
\end{figure*}

\subsection{Prompt for Zero-Shot Model}

For models like MiDashengLM-7B that are evaluated directly without fine-tuning, we carefully design the prompt templates shown below to maximize the model's performance while ensuring a fair comparison.

For the four speech quality assessment tasks, we developed dedicated prompt templates to guide the model in performing precise, context-aware evaluations.

Speech Quality Assessment focuses on analyzing a single speech sample in terms of clarity, naturalness, emotional expression, and overall listening experience, as illustrated by the prompt template in Figure~\ref{fig:sqa-prompt}.

Speech Quality Comparison compares two samples, highlighting differences in intelligibility, distortion, dynamic range, and artistic expression to judge which sample performs better. The corresponding prompt template is shown in Figure~\ref{fig:sqc-prompt}.

Speech Quality Improvement Suggestion provides actionable recommendations for enhancing a speech sample, addressing issues such as pacing, pronunciation, emotional conveyance, and technical quality, with the prompt presented in Figure~\ref{fig:sqi-prompt}.

Deepfake Speech Detection determines whether a speech sample is human-produced or synthetically generated, offering a binary real/fake judgment.  The prompt diagram is depicted in Figure~\ref{fig:dsd-prompt}.

Each prompt outlines evaluation criteria and output requirements, and attaches the relevant audio file paths (two for the SQC task and one for the others). Following standardization, we performed inference on these four tasks using three baseline models from the ms-swift framework.

\begin{table}[t]
\centering
\resizebox{\linewidth}{!}{%
{%
\renewcommand{\arraystretch}{1.15}%
\begin{tabular}{lrrrr}
\hline
\multicolumn{1}{l}{\bf Lang} &
\multicolumn{1}{c}{\bf Real} &
\multicolumn{1}{c}{\bf Open} &
\multicolumn{1}{c}{\bf Comm.} &
\multicolumn{1}{c}{\bf Overall} \\
\hline
English  & 93.3  & 85.9  & 97.9  & 88.6 \\
French   & 100.0 & 99.9  & 53.5 & 86.6 \\
Japanese & 99.9  & 100.0 & 95.9  & 99.1 \\
Chinese  & 99.9  & 69.9  & 97.0  & 82.9 \\
\hline
\end{tabular}%
}%

}
\caption{Deepfake detection accuracy (\%) by language, broken down by real speech and two fake-source types (open-source and commercial), together with the overall results.}
\label{tab:real_fake_by_lang}
\end{table}

\section{Additional Results}
\label{appendixresults}

\paragraph{Speech Quality Assessment}
Table \ref{tab:singleeval_pcc} presents dimension-wise Pearson correlations with human ratings and speech rate accuracy. In the direct-evaluation setting (Qwen2-Audio-7B, Qwen2.5-Omni-7B, MiDashengLM-7B), models generally struggle to capture speech quality; Qwen2.5-Omni-7B is notably strong on speech rate, achieving the best SR ACC without task-specific training, indicating that the pretrained model has a strong grasp of speech tempo. When we move to trained systems, custom-constructed pipelines, or fine-tuned audio LLMs raise several dimensions, yet the gains are uneven across categories. In contrast, SQ-LLM delivers uniform improvements in PCC on every dimension and yields the strongest average correlation with human raters. This pattern is consistent with the effect of CoT guidance combined with GRPO, which helps the model distinguish low-level degradations from higher-level prosodic and affective cues and reason about them explicitly.

\paragraph{Deepfake Detection Analysis}
\label{app:deepfake}

Table~\ref{tab:real_fake_by_lang} further breaks down deepfake detection accuracy by language and fake-source type. Overall performance is strong on Japanese, with consistently high accuracy across all categories. In contrast, the overall scores drop for Chinese and French, but the degradation is source-dependent: Chinese is mainly affected by open-source fakes , while French shows a sharp decline on commercial fakes. Notably, accuracy on real speech remains near-perfect across languages, suggesting low false-positive rates; the main weakness lies in generalizing to specific fake generation pipelines.

\paragraph{CoT Reasoning Analysis}
Figure~\ref{fig:cot_pcc} presents a comparative analysis of SQ-LLM models trained with and without Chain-of-Thought reasoning. The results show that while both models perform comparably on overall quality and distortion, CoT reasoning leads to noticeable improvements on more subjective and perceptual aspects. Specifically, dimensions like dynamic range, emotional impact, artistic expression, and dynamic range benefit the most from CoT supervision, indicating that step-by-step reasoning helps the model better capture subtle cues in expressiveness and listener experience.

\end{document}